\documentclass[aps,prd,floatfix,superscriptaddress,preprintnumbers,nofootinbib,preprint,tightenlines,longbibliography]{revtex4-1}
\pdfoutput=1
\usepackage[english]{babel}
\usepackage{amsmath,amssymb,amsbsy,amstext,amsthm,booktabs,exscale,relsize,slashed,graphicx,amsfonts,upgreek,xcolor,afterpage}
\usepackage{multirow,dcolumn,bm,enumerate,url,hyperref}
\definecolor{lightsaberblue}{rgb}{.0,.1,.5}

\hypersetup{
  colorlinks   = true, %Colors links instead of ugly boxes
  urlcolor     = lightsaberblue, %Color for external hyperlinks
  linkcolor    = lightsaberblue, %Color of internal links
  citecolor   = lightsaberblue %Color of citations
}

\newcommand{\gev}{{\rm GeV}}
\newcommand{\mev}{{\rm MeV}}

\begin{document}
\title{\bf Diffuse X-Ray and Gamma-Ray Limits on Boson Stars \\ that Interact with Nuclei}

\author{Javier F.  Acevedo}
\thanks{{\scriptsize Email}: \href{mailto:17jfa1@queensu.ca}{17jfa1@queensu.ca}; {\scriptsize ORCID}: \href{http://orcid.org/0000-0003-3666-0951}{0000-0003-3666-0951}}
\affiliation{\smaller The Arthur B. McDonald Canadian Astroparticle Physics Research Institute, \protect\\ Department of Physics, Engineering Physics, and Astronomy, \protect\\ Queen's University, Kingston, ON K7L 2S8, Canada}

\author{Amit Bhoonah}
\thanks{{\scriptsize Email}: \href{mailto:Amit.Bhoonah@colostate.edu}{Amit.Bhoonah@colostate.edu}; {\scriptsize ORCID}: \href{https://orcid.org/0000-0002-4206-215X}{0000-0002-4206-215X}}
\affiliation{\smaller Department of Physics, Colorado State University, Fort Collins, CO 80523, USA}

\author{Joseph Bramante}
\thanks{{\scriptsize Email}: \href{mailto:joseph.bramante@queensu.ca}{joseph.bramante@queensu.ca}; {\scriptsize ORCID}: \href{http://orcid.org/0000-0001-8905-1960}{0000-0001-8905-1960}}
\affiliation{\smaller The Arthur B. McDonald Canadian Astroparticle Physics Research Institute, \protect\\ Department of Physics, Engineering Physics, and Astronomy, \protect\\ Queen's University, Kingston, ON K7L 2S8, Canada}
\affiliation{\smaller Perimeter Institute for Theoretical Physics, Waterloo, ON N2J 2W9, Canada}

\begin{abstract}
Light bosonic dark matter can form gravitationally bound states known as boson stars. 
In this work, we explore a new signature of these objects interacting with the interstellar medium (ISM). 
We show how small effective couplings between the bosonic dark matter and the nucleon lead to a potential that accelerates ISM baryons as they transit the boson star, making the ISM within radiate at a high rate and energy. 
The low ISM density, however, implies the majority of Galactic boson stars will be too faint to be observable through this effect. 
By contrast, the diffuse photon flux, in hard x-rays and soft gamma-rays, produced by boson stars interacting with the ionized ISM phases can be sizable. 
We compute this diffuse flux and compare it to existing observations from HEAO-1, INTEGRAL and COMPTEL to infer limits on the fraction of these objects. 
This novel method places constraints on boson star dark matter while avoiding back-action effects from ambient baryons on the boson star configuration, unlike terrestrial searches where it has been noted that back-action can screen light bosonic fields. In addition, this study could be extended to other couplings and structures formed from light dark matter. 
For dark matter masses $(10^{-14}$, $10^{-8}) \ {\rm eV}$ and boson star masses $(10^{-10}$, $10^{-1}) \ M_{\odot}$, we find the constraints on the fraction can go down to $f_* \lesssim 10^{-9}$ for dark matter in boson stars that is directly coupled to the Standard Model. 
\end{abstract}
\maketitle

\newpage
\tableofcontents

%\newpage
\section{Introduction}
\label{sec:int}
The existence of dark matter (DM) is well-established from its gravitational interactions with visible matter. Yet, despite its large abundance relative to its visible counterpart, the origin and nature of dark matter remain a mystery for modern science. It is possible that dark matter has couplings to the visible sector besides its observed gravitational interactions. A myriad of dark matter models have been proposed, and numerous collaborations are searching for their signatures in diverse settings, including underground detectors, particle colliders and astronomical observatories, with even more sensitive experiments proposed for the coming decades. 

Much like visible matter in our universe, dark matter could reside in composite objects. Composite dark matter models include fermionic dark matter composites \cite{Krnjaic:2014xza,Wise:2014ola,Gresham:2017zqi,Gresham:2017cvl,Grabowska:2018lnd,Acevedo:2020avd,Acevedo:2021kly}, glueballs \cite{Soni:2016gzf}, dark quark nuggets \cite{Zhitnitsky:2002qa,Bai:2018dxf}, dark baryons \cite{Cline:2022leq}, massive compact mini-halos \cite{Ricotti:2009bs,Bai:2020jfm}, bound axionic states \cite{Barranco:2010ib,Guth:2014hsa,Eby:2014fya,Braaten:2015eeu,Bai:2016wpg,Braaten:2016dlp,Visinelli:2017ooc,JacksonKimball:2017qgk} and boson stars \cite{Feinblum:1968nwc,Baldeschi:1983mq,Colpi:1986ye,Lee:1995af,Sahni:1999qe,Zurek:2006sy,Boehmer:2007um,Chavanis:2011zi,Liebling:2012fv,Eby:2015hsq,Gorghetto:2022sue,Amin:2022pzv,Jain:2022nqu,March-Russell:2022zll,Collier:2022cpr,Cardoso:2022vpj,Chan:2022bkz,Sanchis-Gual:2022zsr}. In fact, some of these models have lately become the special target of a number of experimental efforts \cite{Bramante:2018qbc,Bramante:2018tos,Bramante:2019yss,Acevedo:2020gro,Acevedo:2020avd,Acevedo:2021kly,Carney:2019pza,Cappiello:2020lbk,Ebadi:2021cte,Acevedo:2021tbl,Ebadi:2021cte,Bhoonah:2020fys,DEAPCollaboration:2021raj,Carney:2022gse}. While the lightest dark matter candidates can be directly searched for in terrestrial experiments, flux limitations impede conventional terrestrial detectors from finding the heaviest dark matter candidates, since the dark matter flux expected at Earth is around one Planck mass of dark matter per year per square meter. As a consequence, it is usually more promising to seek out macroscopic and heavy composite dark matter theories through astrophysical or cosmological signatures, including accretion \cite{Bai:2020jfm}, gravitational lensing \cite{Katz:2019qug,Bai:2020jfm,Croon:2020wpr,Croon:2020ouk}, the disruption of stellar evolution \cite{Acevedo:2020gro,Das:2021drz,Bramante:2021dyx,Dessert:2021wjx} and the alteration of various cosmological observables \cite{Cyr-Racine:2012tfp,Cline:2013zca,Gluscevic:2017ywp,Gresham:2018anj,Ponton:2019hux,Hong:2020est,Mathur:2021gej}. 

In particular, boson stars have long been an appealing astrophysical object, considering the ubiquity of light bosonic degrees of freedom in theories beyond the Standard Model like axions \cite{Peccei:1977hh,Peccei:1977ur,Weinberg:1977ma,Dine:1981rt} and axion-like particles \cite{Arvanitaki:2009fg,Graham:2015cka}, moduli \cite{Seiberg:1994aj,Seiberg:1994rs} and dilatons \cite{Damour:1994zq}, dark photons \cite{Holdom:1985ag,Cvetic:1995rj}, fuzzy dark matter \cite{Hu:2000ke} and quintessence \cite{Ratra:1987rm}. In essence, boson stars are large collections of light bosons in a gravitationally-bound configuration, which at low temperature becomes a self-gravitating Bose-Einstein condensate. The condensed phase is described by a classical scalar field, overlapped with quantum fluctuations that account for the collective modes in the condensate. Various numerical simulations indicate these objects could have formed through several mechanisms within dark matter halos during structure formation early on \cite{Schive:2014hza,Schwabe:2016rze,Levkov:2018kau,Widdicombe:2018oeo,Amin:2019ums,Chen:2020cef}. In the case of Newtonian boson stars, the gravitational field sourced by these objects can be weak compared to most stellar objects, so detecting or constraining these objects by searching for purely-gravitational processes like accretion or lensing can be challenging. By contrast, the bosonic field they source attains a large enough expectation value in their interior that Standard Model particles, if coupled to this field, accelerate to significant energies as they pass through. This effect is appreciable even for couplings between the bosonic field and the Standard Model below existing experimental limits. Indeed, in this work we see that boson stars with relatively straightforward couplings to the Standard Model can be highly efficient at heating up matter via the resulting scalar potential. 

There are various astrophysical signatures which could arise from this effective potential sourced by the boson stars. In this work, we will focus on a robust signal that arises from boson stars steadily accelerating baryons from the interstellar medium (ISM). Because of the low ISM density, potential screening effects of the ambient matter on the scalar field configuration, the so-called back-action, are negligible for most of the parameter space. As has been noted in prior works, back-action could severely limit the detection prospects of light dark matter at experiments near the Earth's surface \cite{Hinterbichler:2010es,Elder:2016yxm,Hees:2018fpg,Stadnik:2020bfk,Masia-Roig:2022net}. By modeling the ISM matter flow as nearly collisionless, suitable for small Newtonian boson stars within the most rarefied ISM phases, we have found that the baryons accelerated within the boson star internal potential will radiate photons at a greater rate than the unperturbed ISM. For the parameter space we will be interested in, the radiation emitted by the baryons within boson stars is insufficient to detect a boson star as a point source in the ISM. Instead, we find that the collective emission of all Galactic boson stars in the ISM results in a substantial and detectable diffuse photon flux. When compared with data from existing diffuse photon observations, this flux constrains the fraction of light dark matter that can be contained in boson stars and also be weakly-coupled to the Standard Model. In particular, we will focus here on diffuse emission of hard x-ray and soft gamma-ray components, an energy range in which extensive searches for dark matter annihilation and decay have been carried out \cite{Boehm:2003bt,Huh:2007zw,Hooper:2008im,Khalil:2008kp,Goodenough:2009gk,Vincent:2012an,Essig:2013goa,Bulbul:2014sua,Boyarsky:2014ska,Daylan:2014rsa,Lee:2014mza,Iakubovskyi:2014yxa,Cline:2014vsa,Hofmann:2016urz,Dessert:2018qih,Ng:2019gch,Leane:2019xiy,Leane:2020nmi,Abazajian:2020tww,Boyarsky:2020hqb,Coogan:2020tuf,Laha:2020ivk}. To our knowledge, this is the first such indirect detection method proposed for this class of macroscopic dark matter objects. For fixed dark matter-nucleon couplings, we will obtain constraints on the boson star fraction by analyzing the various radiative processes that occur from ISM baryons collected in boson stars, and comparing the resulting diffuse photon flux to existing data from the High Energy Astronomy Observatory (HEAO), the International Gamma Ray Astrophysics Laboratory (INTEGRAL), and the Imaging Compton Telescope (COMPTEL). By summing the contributions of all the Galactic boson stars located in the large galactic volumes inhabited by ionized ISM, stringent constraints can be placed on the boson star fraction of dark matter for dark matter-baryon couplings below existing experimental limits. Such constraints apply to boson masses $(10^{-14}$, $10^{-8}) \ {\rm eV}$, boson star masses $(10^{-10}$, $10^{-1}) \ M_{\odot}$ and repulsive boson self-couplings up to $\lesssim 10^{-52}$. This novel method allows us to probe light dark matter interactions if a sizable fraction of the dark matter content has formed compact structures that rarely intersect with the Earth, therefore limiting detection prospects at terrestrial experiments. 

The outline of this paper is as follows: 
in Section~\ref{sec:bssol} we review the set of semi-analytic solutions for spherically symmetric Newtonian boson stars that we utilize in this work, for both weak and strong boson self-interactions. 
Section~\ref{sec:radflux} is divided into the following: we first show in Subsection~\ref{subsec:bspotential} how simple effective couplings between SM nucleons and the bosonic field lead to an effective potential under which baryons from the ISM can accelerate to significant energies. We also discuss constraints on effective couplings and potential back-action effects from ISM matter on the bosonic field configuration. We show in Subsection~\ref{subsec:accretesol} how the ISM density and temperature within boson stars can be reasonably obtained assuming a collisionless flow of particles, as long as the boson star is small enough and the background ISM is sufficiently rarefied. 
Next, in Subsection~\ref{subsec:radiative}, we analyze the emission of final state radiation that proceeds in the heated ISM matter within boson stars. 
Section~\ref{sec:constraints} is divided as follows: we first compute the luminosity of boson stars in the ISM and its relevant scalings in Subsection~\ref{subsec:lum}. Next, we compute the diffuse photon flux that results from adding the contribution of all boson stars along a certain line of sight in Subsection~\ref{subsec:diff}. 
Based on existing observations, we derive limits on the fraction of dark matter within boson stars for various fixed effective couplings and dark matter masses in Subsection~\ref{subsec:limits}.
Finally, in Section~\ref{sec:concl} we provide a brief summary of our results and conclude. Throughout this paper, we use natural units where $c = \hbar = k_b =1$ and $G = M_{\rm pl}^{-2}$, where $M_{\rm pl} \simeq 1.2 \cdot 10^{19} \ \gev$ is the Planck mass.

\section{Boson Star Solutions}
\label{sec:bssol}
Field solutions for bound systems of self-interacting bosons have been investigated in a number of past references \cite{Kaup:1968zz,Feinblum:1968nwc,Ruffini:1969qy,Colpi:1986ye,Agnihotri:2008zf,Barranco:2010ib,Chavanis:2011zi,Chavanis:2011zm,Eby:2014fya,Eby:2015hsq,Kling:2017mif,Kling:2017hjm}. Here we briefly review the semi-analytical solutions obtained in \cite{Kling:2017hjm,Kling:2017mif}, which are utilized in this work. These solutions are consistent with earlier numerical approaches \cite{Chavanis:2011zi,Chavanis:2011zm}, yet they are less computationally expensive. We start from the scalar field theory in the presence of gravity,
\begin{equation}
    \mathcal{L} = \frac{1}{2} g^{\mu \nu} \partial_{\mu} \phi^\dagger \partial_{\nu} \phi - \frac{1}{2} m^2 \phi^{\dagger} \phi - \frac{\lambda}{4!} \left(\phi^{\dagger} \phi\right)^2,
\end{equation}
where the scalar field represents the bosonic dark matter and can either be complex or real. The self-interaction between bosons is attractive when $\lambda < 0$ and repulsive when $\lambda > 0$. In the case of attractive self-interactions, the potential is unbounded from below, and this requires the inclusion of higher-dimensional operators to stabilize the field. These higher-order operators have no impact on the phenomenology outlined here as long as their cutoff scale is $\gg m/\sqrt{|\lambda|}$. We assume the boson star to be at zero-temperature, so that all the bosons are in their ground state and no thermally excited collective modes are occupied. Moreover, following \cite{Kling:2017mif,Kling:2017hjm} we assume that:
\begin{enumerate}
    \item the boson star is non-relativistic, $i.e.$ the energy of a particle is expressed as $E=m+b$ where $b \ll m$ is its binding energy, and
    \item the boson star is light enough that general relativistic corrections are negligible, $i.e.$ we consider Newtonian boson stars. The inverse metric $g^{\mu\nu}$ is then expanded as ${\rm diag}\left((1+2\Phi)^{-1},-1,-1,-1\right)$, where $\Phi \ll 1$ is the Newtonian gravitational potential. 
\end{enumerate}
With these considerations, the scalar field can be expanded as
\begin{equation}
\phi(r,t) =
    \begin{cases}
      \ \left(\frac{2m^2}{M_*}\right)^{-1/2} \psi(r) \exp(-imt)  & \text{(Complex Scalar)} \\
      \\
      \ \left(\frac{2m^2}{M_*}\right)^{-1/2} \psi(r)\left(\exp(-imt)+\exp(imt)\right) & \text{(Real Scalar)}\\
    \end{cases}
    \label{eq:phiexp}
\end{equation}
where $M_* \simeq N m$ is the boson star mass and $N$ is the number of bosons. The field $\psi(r)$ is real and normalized so that $\int \psi^2 dV = 1$, and is subsequently interpreted as a probability amplitude. The equations of motion for this field are the Gross-Pitaevskii + Poisson (GPP) equations
\begin{equation}
    -\frac{1}{2m}\nabla^2 \psi + m \Phi \psi + \frac{M_* \lambda}{4 m^3} \psi^3 = m \psi
    \label{eq:GPP1}
\end{equation}
\begin{equation}
    \nabla^2 \Phi = 4 \pi G M_* \psi^2
    \label{eq:GPP2}
\end{equation}
with boundary conditions $\psi'(r=0)=0$, $\psi(r\rightarrow\infty)\rightarrow 0$ and $\Phi(r\rightarrow\infty) \rightarrow b/2m$. The left hand side terms of Eq.~\eqref{eq:GPP1} are respectively the quantum pressure, gravity and classical pressure contributions to the energy of a boson in this bound state. The solution to these equations is highly sensitive to the boson self-interaction sign and strength. For what follows, we introduce the dimensionless parameter defined in \cite{Kling:2017hjm},
\begin{equation}
    \gamma = \frac{b \lambda}{8\pi G m^3}~,
    \label{eq:gamma}
\end{equation}
which is a measure of the self-interaction strength. 
The range $-1 \lesssim \gamma \lesssim 1$, corresponds to weak self-interactions, with the sign determining whether they are repulsive or attractive. In practice, there is a lower bound of $\gamma \gtrsim -0.72$ below which no stable static solution for a boson star is found \cite{Kling:2017hjm}. On the other hand, $\gamma \gtrsim 1$ corresponds to strong repulsive self-interactions. Limiting values of the binding energy per particle are discussed below for each case. 

\subsection{Weak Self-Interactions}
In the case of weak self-interactions, the quantum pressure balances the self-gravity of the bosons and the classical pressure can be treated as a perturbation. In this limit, the binding energy per boson scales as $b \sim G^2 M_*^2 m^3$, and the self-interaction parameter $\gamma$ therefore reads
\begin{equation}
    \gamma \simeq \frac{GM_*^2\lambda}{8\pi}~.
    \label{eq:gammaweak}
\end{equation}

For $\gamma$ values in the range $|\gamma|\lesssim 1$, we find no significant variation of our results compared to the case without self-interactions, $i.e.$ $\gamma = 0$. Therefore, we will review mostly the solution without self-interactions, and discuss below how weak self-interactions can be introduced as a perturbation to the solution. Ref.~\cite{Kling:2017mif} has found a semi-analytic expression for the profile that approximates the full numerical solution of Eqs.~\eqref{eq:GPP1} and \eqref{eq:GPP2} with sub-percent accuracy. In terms, of the auxiliary field $\psi(r)$, this is 
\begin{equation}
    \psi(r) \simeq \frac{1}{\sqrt{8\pi}} \left(\frac{G^{3/2} M_*^{3/2} m^3}{\beta^2}\right) s\left(\frac{G M_* m^2 }{\beta} r \right)~,
    \label{eq:psi0}
\end{equation}
where the function $s(z)$ is
\begin{equation}
    s(z)\simeq
    \begin{cases}
      \ s_0 - 0.159\ z^2 +1.63 \cdot 10^{-2}\ z^4 + \mathcal{O}(z^6)~,  & z < 2.5 \\
      \\
      \ z^{0.7526}\exp(-z)\left(3.4951-2.3053\ z^{-1}\right)~, & z > 2.5\\
    \end{cases}
\end{equation}
where $s_0 \simeq 1.0214$ and $\beta \simeq 1.7527$ is the so-called mass parameter. These coefficients are obtained from the full numerical solution of Eqs.~\eqref{eq:GPP1} and \eqref{eq:GPP2}, which are utilized as expansion parameters for the power series solution at small distance from the boson star center, and in turn matched to the power series solution at large distances \cite{Kling:2017mif}. Note that the field rapidly decays past the (dimensionless) radial distance $z_* \gtrsim 2.5$, and so we will take the boson star radius in this case to be approximately
\begin{equation}
    R_* \simeq \left(\frac{G M_* m^2}{\beta}\right)^{-1} \simeq 7 \ R_{\odot} \left(\frac{m}{10^{-8} \ \rm eV}\right)^{-2} \left(\frac{M_*}{10^{-10} \ M_\odot}\right)^{-1}~.
    \label{eq:bsradius1}
\end{equation}
Comparing this radial scale to the boson star mass, we see that these objects source gravitational fields that are weaker than most stellar objects when the total mass is much smaller than the maximum threshold,
\begin{equation}
    M_{*}^{\rm max} \simeq \frac{M_{\rm pl}^2}{m} \simeq 10^{-2} \ M_{\odot} \left(\frac{m}{10^{-8} \ \rm eV}\right)^{-1}.
\end{equation}
In practice, this will always be the case for the parameter space we consider here.

\begin{figure}[h!]
\includegraphics[width=0.8\textwidth]{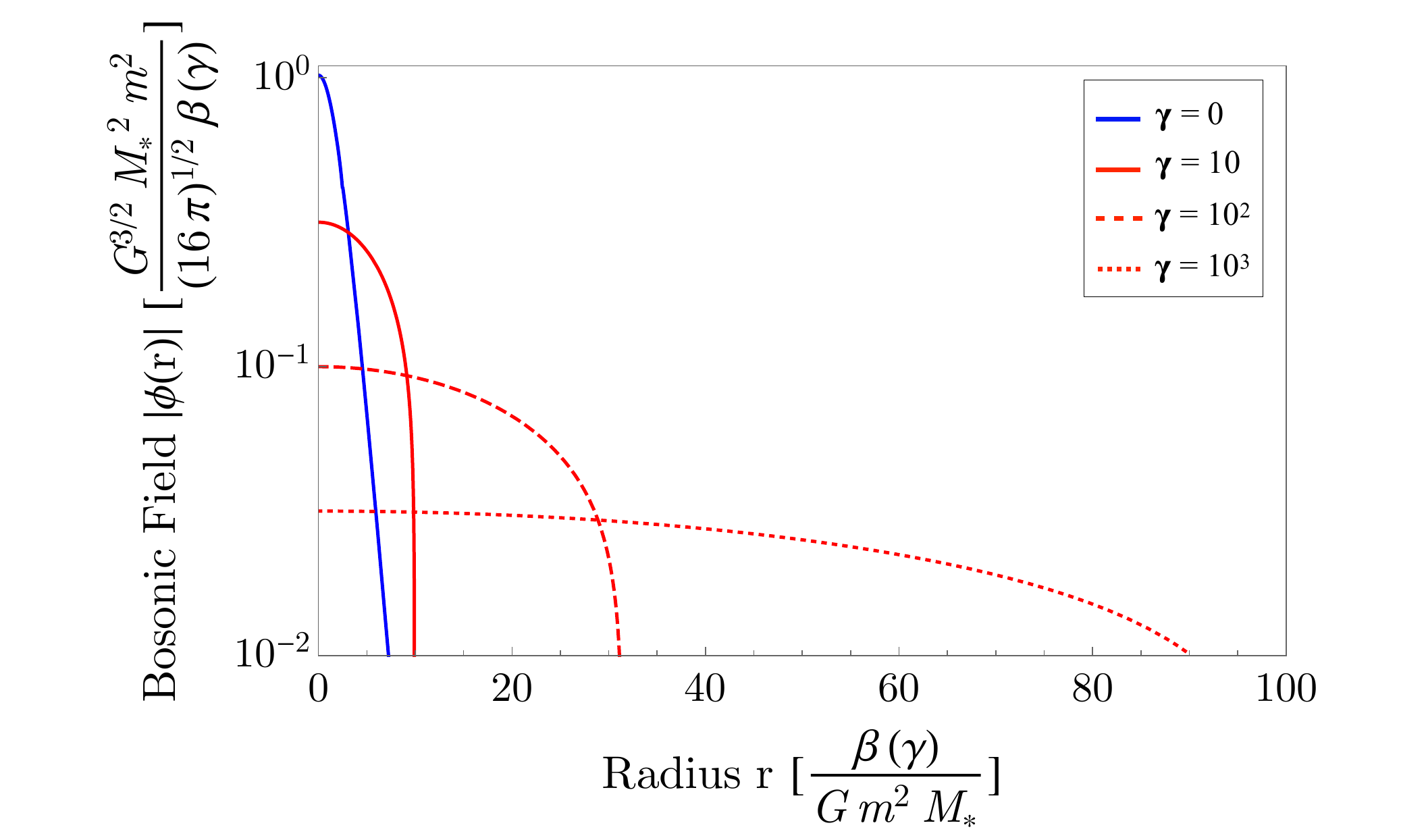}
\caption{Boson star field amplitude profile as a function of boson star radius, obtained using the semi-analytical solutions of \cite{Kling:2017hjm,Kling:2017mif}, for various self-interaction strengths $\gamma$ as specified, $cf.$ Eqs.~\eqref{eq:psi0} and \eqref{eq:psiTF}. The radius has been scaled in units of $\beta(\gamma) /(Gm^2M_*)$, since this quantity describes the radius beyond which the field profile truncates, $cf.$ Eq.~\eqref{eq:psiTF}.}
\label{fig:bsprof0}
\end{figure}

For weak but finite boson self-interactions, the effects of the classical pressure can be incorporated into the above solution as a perturbation to its various parameters. In particular, the mass parameter and the central field value are now corrected by \cite{Kling:2017hjm}
\begin{equation}
    \beta \rightarrow \beta + 0.7039 \ \gamma +\mathcal{O}(\gamma^2)~,
\end{equation}
\begin{equation}
    s_0 \rightarrow s_0-0.3909\ \gamma+\mathcal{O}(\gamma^2)~.
\end{equation}
Eq.~\eqref{eq:psi0}, along with the appropriate expression for $\phi$ in Eq.~\eqref{eq:phiexp} and the corrections to the parameters that account for weak self-interactions, is the semi-analytic solution. The resulting field amplitude profile $|\phi(r)|$ is illustrated in Figure~\ref{fig:bsprof0} for $\gamma = 0$.

\subsection{Strong Self-Interactions}
The strong repulsive self-interaction limit corresponds to the case where the classical pressure balances the self-gravity of the bosons, and the quantum pressure becomes sub-dominant. This is the so-called Thomas-Fermi limit, which requires the self-interaction parameter $\gamma$ to be much larger than unity. The binding energy scales as $b \sim \lambda^{-1/2} G^{3/2} m^3 M_*$, and therefore
\begin{equation}
    \gamma \simeq \left(\frac{\lambda G M_*^2}{4 \pi^3}\right)^{1/2}.
    \label{eq:gammaTF}
\end{equation}

Similarly to the previous case, the auxiliary field $\psi(r)$ is expressed as
\begin{equation}
    \psi(r) \simeq \frac{1}{\sqrt{8 \pi}} \left(\frac{G^{3/2} M_*^{3/2} m^3}{\beta(\gamma)}\right) s\left(\frac{G M_* m^2 }{\beta(\gamma)} r \right)~,
    \label{eq:psiTF}
\end{equation}
however an analytic solution to the boson star profile can be recovered when fully neglecting the quantum pressure, and is given by \cite{Slepian:2011ev}
\begin{equation}
    s(z) = s_0(\gamma) \ {\rm sinc}^{1/2} \left(\gamma^{-1/2} z\right)~,
    \label{eq:sinc}
\end{equation}
where now $\beta (\gamma) = (\pi/2) \gamma^{1/2}$ and $s_0(\gamma) = \gamma^{-1/2}$. The effects of a finite quantum pressure are then introduced as a correction to the mass parameter and the central field value \cite{Kling:2017hjm}
\begin{equation}
    \beta(\gamma) \rightarrow \beta(\gamma) \left[1-0.001478 \  \gamma^{-1/3}+\mathcal{O}\left(\gamma^{-2/3}\right)\right]~,
\end{equation}
\begin{equation}
    s_0(\gamma) \rightarrow s_0(\gamma) \left[1+0.003712 \ \gamma^{-1/2} + \mathcal{O}\left(\gamma^{-2/3}\right)\right]~.
    \label{eq:s(r)TF}
\end{equation}
Note that the solution given by Eq.~\eqref{eq:sinc} implies the boson star profile now has a compact support out to dimensionless radius $z_* = \gamma^{1/2} \pi$. In terms of the basic boson star parameters, this is
\begin{equation}
    R_* \simeq 210 \ R_{\odot} \left(\frac{\gamma}{10}\right) \left(\frac{m}{10^{-8} \ \rm eV}\right)^{-2} \left(\frac{M}{10^{-10} \ M_{\odot}}\right)^{-1}.
    \label{eq:bsradius2}
\end{equation}
It can be immediately seen that boson stars with repulsive self-interaction have much more extended configurations. In particular, the maximum stable mass is significantly increased in this regime,
\begin{equation}
    M_*^{\rm max} \simeq \sqrt{\lambda} \frac{M_{pl}^3}{m^2} \simeq 10^8 \ M_{\odot} \left(\frac{\lambda}{10^{-52}}\right)^{1/2} \left(\frac{m}{10^{-8} \ \rm eV}\right)^{-2}.
\end{equation}
Eq.~\eqref{eq:psiTF}, combined with the appropriate expression for $\phi$ in Eq.~\eqref{eq:phiexp} and the corrections to the parameters for a finite $\gamma$, is the semi-analytic solution. Field amplitude profiles $|\phi(r)|$ in this limit are also shown in Figure~\ref{fig:bsprof0} for various self-interaction strengths.

\section{Radiation from Boson Stars Interacting with the ISM}
\label{sec:radflux}

\subsection{Boson Star Effective Potential for ISM baryons}
\label{subsec:bspotential}
Using the previously outlined solutions for boson stars, we now show how these objects can accelerate ISM particles to sizable energies when a small effective coupling to the nucleon field is introduced. We consider here a quadratic interaction between the dark matter and the Standard Model nucleon of the form
\begin{equation}
    \mathcal{L}_{\rm int} = + \left(\frac{m_n}{\Lambda^2}\right) n\bar{n} \phi \phi^{\dagger},
    \label{eq:nongcoupling}
\end{equation}
although other similar effective couplings to the nucleon could be realized, see $e.g.$~\cite{Brax:2017xho}. The motivation for the sign choice of Eq.~\eqref{eq:nongcoupling} is detailed below. Quadratic portals to light scalar dark matter have been considered in many recent works \cite{Olive:2007aj,Derevianko:2013oaa,Stadnik:2014tta,Arvanitaki:2014faa,VanTilburg:2015oza,Stadnik:2015kia,Stadnik:2015xbn,Elder:2016yxm,Kalaydzhyan:2017jtv,Hees:2018fpg,Dailey:2020sxa,Masia-Roig:2022net,Berger:2022tsn}, most of which are focused on the terrestrial detection of ultralight bosonic dark matter utilizing precise atomic systems. These studies typically assume that most of the dark matter density is in the form of coherent waves or macroscopic topological defects, and have placed limits on its interactions based on the field profile near the Earth. In fact, these constraints are not limited to effective nucleon couplings. The phenomenology and detection prospects of similar effective interactions to light quarks, electrons, photons and massive vector bosons have also been analyzed. For simplicity, we will limit our attention to effective couplings to the nucleon field. 

There exist several constraints on the effective coupling scale $\Lambda$ in Eq.~\eqref{eq:nongcoupling} that are derived from experimental, astrophysical and cosmological considerations. If all the dark matter is contained in boson stars ($e.g.$ due to self-interactions or gravitational instabilities), the mean encounter rate of these objects with the solar system roughly scales as $t_{\rm enc} \gtrsim 100 \ {\rm yr} \ (M_{*}/10^{-10} \ M_{\odot})$. In this case, the most robust laboratory constraints on $\Lambda$ come from the E\"{o}t-Wash experiment, a torsion balance experiment primarily dedicated to search for Yukawa-type deviations of the inverse-square law of the gravitational force \cite{Lee:2020zjt}. Effective operators such as Eq.~\eqref{eq:nongcoupling} yield power-law potentials of the form $\propto r^{-n}$ derived from nucleon-nucleon scattering through two-scalar exchange, where $n=3$ corresponds to the above case. The E\"{o}t-Wash experiment has previously placed constraints on such power-law interactions \cite{Adelberger:2006dh}. Following the procedure outlined in \cite{Brax:2017xho}, we map the constraints from \cite{Adelberger:2006dh} into a lower bound for $\Lambda$, which reads $\Lambda \gtrsim 1.3 \ \rm TeV$. On the other hand, stellar cooling places stronger constraints on this cutoff scale, of order $\Lambda \gtrsim 15 \ \rm TeV$, based on the energy loss to light scalar emission through nucleon bremmstrahlung $N N \rightarrow N N \phi \phi$ \cite{Olive:2007aj}. In addition, if the light scalar is produced non-thermally, and its mass is in excess of $m \gtrsim 10^{-16} \ \rm eV$, it can form a classical field that coherently oscillates in the early universe, leading to temporal variations of fundamental physical constants through its couplings to other Standard Model fields \cite{Stadnik:2015kia}. In particular, effective couplings to light quarks alter the neutron decay rate during Big Bang Nucleosynthesis, and therefore the primordial abundance of $^4$He. This places a much stronger constraint of the order $\Lambda \gtrsim 10^{13} \ {\rm TeV} \left(m/10^{-16} \ {\rm eV}\right)^{-1}$. As we also discuss below, linear couplings on the light boson are severely constrained by experiments.

The astrophysical and cosmological constraints on $\Lambda$, albeit stronger than their experimental counterpart, are less robust. Stellar cooling bounds are particularly sensitive to core temperature and therefore subject to its uncertainty, and may also be circumvented with minimal model extensions \cite{DeRocco:2020xdt}.
Similarly, cosmological bounds ultimately depend on the production history of the dark matter, and can also be evaded in alternative cosmological scenarios, see $e.g.$ \cite{Bernal:2021bbv}. In addition, cosmological bounds depend on the strength of the coupling in Eq.~\eqref{eq:nongcoupling} at early times, which can shift due to phase transitions and rolling scalar fields in the early universe, see $e.g.$ \cite{Zhao:2017wmo,Davoudiasl:2019xeb,Cohen:2008nb,Bhoonah:2020oov} for model building along these lines.  
Hence for robustness, from hereon we will consider effective coupling scales above the current experimental limits. Specifically, we will discuss our results for the benchmark value $\Lambda \simeq 2 \ \rm TeV$, marginally above the lower experimental bound, as well as $\Lambda \simeq 10 \ \rm TeV$ to illustrate how the photon flux produced by boson stars weakens with smaller effective couplings. As we detail in our results below, coupling scales in the $\Lambda \sim 1$ - $100 \ \rm TeV$ range are suitable for the ISM matter to heat up and emit sizable amounts of x-rays and gamma-rays as it interacts with the boson star internal field, $cf.$ Eqs.~\eqref{eq:ISMaccret1} and \eqref{eq:ISMaccret2}. Smaller effective couplings could in principle be probed by considering lower energy signals, such as radio waves. Since the radio domain involves very different background and attenuation considerations, we have left this for future exploration. We comment that such a study could potentially be applicable to linear couplings in $\phi$, which at present are constrained to values of order $\lesssim 10^{-22}$ by experimental 5th force searches for boson masses $m \lesssim 10^{-3} \ \rm eV$ \cite{Murata:2014nra}. 

At zero temperature, the light scalars form a condensate, and the field $\phi$ is described by a classical field which we obtain from the semi-analytical solutions outlined in Section~\ref{sec:bssol}. By virtue of the effective boson-nucleon coupling given by Eq.~\eqref{eq:nongcoupling}, a nucleon acquires an effective mass of the form
\begin{equation}
 m_n^{\rm eff}(r,t) = m_n \left[1-\left(\frac{|\phi(r,t)|}{\Lambda}\right)^2\right]
\end{equation}
while transiting the boson star. In the limit $|\phi(r,t)| \ll \Lambda$, this mass shift is small, and the kinetic energy change of a nucleon can be expressed in terms of a non-relativistic potential \cite{Acevedo:2020avd,Acevedo:2021kly},
\begin{equation}
    V_{\phi}(r,t) = - m_n \left(\frac{|\phi(r,t)|}{\Lambda}\right)^2~.
    \label{eq:nrpotential}
\end{equation}
For simplicity, we shall only consider the non-relativistic limit in which Eq.~\eqref{eq:nrpotential} is valid. Note that the mass shift depends on the sign of the effective operator written in Eq.~\eqref{eq:nongcoupling}. Here, we will only consider negative mass shifts that yield a boost in kinetic energy as baryons approach the boson star ($i.e.$ boson stars that source attractive potentials), and we will analyze other effective operators including the additive inverse of Eq.~\eqref{eq:nongcoupling} in upcoming work. For a complex field, $cf.$ Eq.~\eqref{eq:phiexp}, the above potential is constant. In what follows, we will show our results for the complex case and therefore omit the time argument, $i.e.$ $V_\phi(r)$. By contrast, for a real scalar, the above potential oscillates with a period set by the inverse boson mass $m^{-1}$. In practice, this oscillation period is short compared to the dynamical crossing time of ISM particles as they transit the boson star. This allows us to average out the time variation of the potential in our treatment and, overall, we have found in preliminary computations that the limits on real scalar bosons star interactions with the ISM (as shown in $e.g.$ Figure \ref{fig:fractionbounds1}) would be weaker by an order $1$-$10$ factor compared to the complex field solution.

Finally, we consider how the effective operator in Eq.~\eqref{eq:nongcoupling} could induce a mass correction $\delta m$ to the bosons in the presence of a background baryon density, the so-called back-action of ambient matter on the scalar field. This is an active topic for detection proposals of light dark matter and associated topological defects in terrestrial and orbital experiments \cite{Hinterbichler:2010es,Elder:2016yxm,Hees:2018fpg,Stadnik:2020bfk,Masia-Roig:2022net}. Parametrically, the effective operator in Eq.~\eqref{eq:nongcoupling} implies a mass correction of order
\begin{equation}
    \delta m^2 \simeq + \frac{\rho_n}{\Lambda^2} \simeq 7\cdot 10^{-32}~{\rm eV^2}~ \left( \frac{10~\rm TeV}{\Lambda}\right)^2 \left( \frac{\rho_n}{1~\rm GeV \ cm^{-3}}\right),
    \label{eq:deltam}
\end{equation}
where $\rho_n \simeq m_n \langle \bar{n}{n} \rangle$ is the baryon energy density in the non-relativistic limit. The preceding sign above is related to the sign of Eq.~\eqref{eq:nongcoupling}, and corresponds to the screening regime which could potentially hamper light dark matter detection in experiments near the Earth's surface. By contrast, in our present work, we avoid back-action effects on the boson star structure for most of the parameter space due to the extremely low density of the ISM. On the other hand, this mass correction could substantially change the cosmological evolution of boson stars for a much larger baryon density in the early universe, however this will depend on the specific dark matter model and formation mechanism for these objects, see $e.g.$ \cite{Schive:2014hza,Schwabe:2016rze,Levkov:2018kau,Widdicombe:2018oeo,Amin:2019ums,Chen:2020cef}.

\subsection{ISM Particle Distribution in the Collisionless Limit}
\label{subsec:accretesol}
The radiation flux produced by the boosted ISM baryons depends on the density and temperature attained within the boson star. For this work, we will focus on boson star parameters for which ISM particles rarely scatter against each other as they transit the boson star. The reason for this requirement is the following: frequent collisions would make the boson star enter a hydrodynamical accretion regime, in which it would rapidly accumulate matter until a static configuration is achieved. The accumulated matter would then cool down on a short timescale through mostly thermal bremsstrahlung, yielding no observable radiation signature. By contrast, if collisions are infrequent, particles accumulate at a low rate and as we show below, the unbound particles output significant radiation over long timescales. Since the potential sourced by the bosonic field is short-ranged, this condition is fulfilled so long as 
\begin{equation}
    2 \sigma n R_* \lesssim 1~,
    \label{eq:mfp-1}
\end{equation}
where $R_*$ is the radius of the boson star, set by Eqs.~\eqref{eq:bsradius1} or \eqref{eq:bsradius2}, $\sigma$ is a scattering cross-section between the ISM particles (see below), and $n$ is the ISM particle number density, which we conservatively set equal to the maximal value we estimate below, $cf.$ Eq.~\eqref{eq:ISMaccret1}. We will only consider boson stars interacting with the warm ionized ISM (WIM) and the hot ISM (HIM), which we assume to be fully ionized. For simplicity, we also assume these phases are entirely composed of electrons and protons. Further details on the WIM and HIM and their structure in the Milky Way are included in Appendix~\ref{sec:app-a}. For elastic scatterings between ISM particles, the appropriate cross-section to use in Eq.~\eqref{eq:mfp-1} is the Coulomb cross-section $\sigma_C$ for elastic scattering of charged particles. At the relevant energies, of order $\rm keV$ and above, this cross-section evaluates to $\sigma_C \lesssim 10^{-22} \ \rm cm^2$. This will be the largest baryon-baryon cross-section for collisions in the boson star, since for sufficiently high-energies of the ISM baryons, proton substructure effects in the scattering become relevant, and the Rosenbluth cross-section must instead be used in Eq.~\eqref{eq:mfp-1}. This cross-section is typically smaller than the cross-section quoted above for all scattering angles \cite{1983MNRAS.202..467S}. In addition, elastic proton-proton scattering proceeds through the strong nuclear force above scattering energies around an MeV. However, in a similar way, the cross-section for this is considerably lower, of order $\sigma_{NN} \simeq 10^{-25} \ \rm cm^2$ \cite{Lechanoine-LeLuc:1993bfx}. The possibility of other inelastic processes, such as isobar production, are detailed in Appendix~\ref{sec:app-e}, but we remark here that they are infrequent as well. Eq.~\eqref{eq:mfp-1} provides a reasonable estimate of the maximum boson star radius $R_*$ for which a collisionless accretion regime can be safely assumed. Due to the overall low density of the WIM and the HIM, we find the condition imposed by Eq.~\eqref{eq:mfp-1} is fulfilled for all the parameter space we consider. In addition to these processes, ISM particles can also scatter against collective modes of the boson star, which are associated with the quantum fluctuations of the field $\phi$ around its expectation value. However, we expect the associated cross-section to be much smaller, of order $\lesssim 10^{-44} \ \rm cm^2$. Further details on this point can be found in Appendix~\ref{sec:app-b}.

\begin{figure}[h!]
\centering 
\centerline{\includegraphics[width=1.15\textwidth]{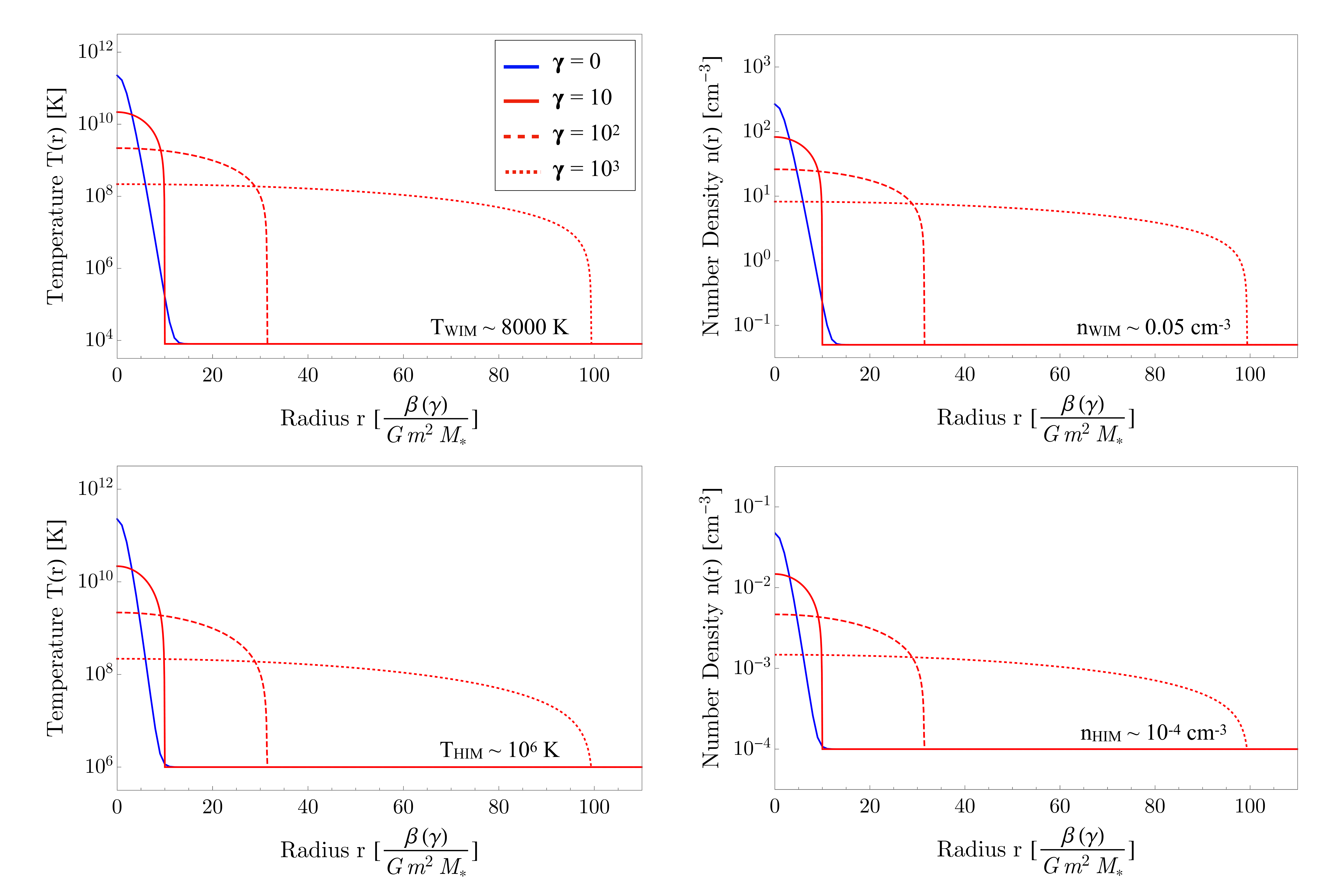}}
\caption{\textbf{Top}: Temperature (\textbf{left}) and number density (\textbf{right}) profiles of ISM particles flowing through a boson star located in the WIM, for different boson self-interaction strengths as specified ($cf.$ Eq.~\eqref{eq:gammaTF}). The boson star parameters are the same as those shown in Figure \ref{fig:bsprof0}. In this plot, we have the effective boson-nucleon coupling scale to $\Lambda \sim 10 \ \left(G^{3/2}M_*^{3/2} m^3 \beta(\gamma)^{-1}\right)$, which means at the center of each boson star the shift in the nucleon mass/potential will be $\sim 10$ MeV $cf.$ Eq.~\eqref{eq:nrpotential}. As is clear this alters the phase-space distribution of the ISM, resulting in an increased ISM density and temperature inside the boson star as compared to the ISM at large distances from the boson star. \textbf{Bottom:} Same as top panels, but for a boson star situated in the HIM.}
\label{fig:ISMprof}
\end{figure}

In the limit that collisions are rare, we approximate the particle flow with a distribution function $f(r,v,t)$ (see $e.g.$~\cite{1983bhwd.book.....S}), which evolves according to 
\begin{equation}
    \frac{Df}{Dt} \equiv \frac{\partial f}{\partial t} + \mathbf{v} \cdot \nabla_{r} f + \mathbf{\dot{v}} \cdot \nabla_v f \simeq 0,
\end{equation}
where the left hand side omits any collisional terms. In the steady state, the distribution function $f$ only depends on the constants of motion $E$ and $J$, which respectively are the total energy and angular momentum of a particle. 
Furthermore, assuming an isotropic flow, the distribution function further simplifies to $f = f(E)$. The specific function $f$ will depend on the distribution function far from the boson star, which we assume to be a Maxwell-Boltzmann distribution $f(E) \simeq n_{\rm ISM} (m/2\pi T_{\rm ISM})^{3/2} \exp(-E/T_{\rm ISM})$. The quantities $n_{\rm ISM}$ and $T_{\rm ISM}$ are the background ISM number density and temperature, and are detailed in Appendix~\ref{sec:app-a} (see also Figure~\ref{fig:ISMprof}). 

The number density and velocity dispersion of unbound particles, $i.e.$ those transiting particles with energies $E>0$, are extracted from the distribution function,
\begin{equation}
    n_{E>0}(r) = \int_{0}^{\infty} f(E) \left[2\left(E-V_{\phi}(r)\right)\right]^{1/2} dE~,
    \label{eq:ISMgen1}
\end{equation}
\begin{equation}
    \langle v^2(r) \rangle_{E>0} = \frac{4 \pi}{n_{E>0}(r)}\int_{0}^{\infty} f(E) \left[2\left(E-V_{\phi}(r)\right)\right]^{3/2} dE~.
    \label{eq:ISMgen2}
\end{equation}
In the limit that boson stars source potentials such that $T_{\rm ISM} \ll |V_{\phi}(r)|$, Eqs.~\eqref{eq:ISMgen1} and \eqref{eq:ISMgen2} can be approximated by
\begin{equation}
    n_{E>0}(r)\simeq n_{\rm ISM} \left(\frac{|V_{\phi}(r)|}{T_{\rm ISM}}\right)^{\frac{1}{2}},
    \label{eq:ISMaccret1}
\end{equation}
\begin{equation}
    T_{E>0}(r)\simeq T_{\rm ISM} \left(\frac{|V_{\phi}(r)|}{T_{\rm ISM}}\right),
    \label{eq:ISMaccret2}
\end{equation}
where $T_{E>0} (r)$ is the temperature of the unbound particles, which we identify as $m_n \langle v^2(r) \rangle_{E>0} \simeq 3 T_{E>0}(r)$. We show how these asymptotic expressions are recovered in Appendix~\ref{sec:app-c}. Figure~\ref{fig:ISMprof} illustrates these temperature and density distributions for various bosonic field profiles, using appropriate boundary values for both the WIM and the HIM. On the other hand, the distribution of bound particles, $i.e.$ those with $E<0$, will depend on the rate at which unbound particles lose energy and become bound to the boson star. This is a significantly more complicated problem which requires analyzing elastic collisions and inverse Compton scattering to accurately determine the accumulation rate. In Appendix~\ref{sec:app-d}, we provide a conservative estimate for the density of ISM particles that accumulate over long timescales due to thermal radiation alone. For most of the parameter space under consideration, we find the density of bound particles is $n_{E<0} \lesssim n_{E>0}$ even if boson stars interact with the ISM on $\sim \rm Gyr$ timescales, implying that our requirements of negligible back-action effects, Eq.~\eqref{eq:deltam}, and long mean free paths, Eq.~\eqref{eq:mfp-1}, are maintained. To be conservative in our limits, we will compute below the diffuse photon flux using the distribution functions for the unbound ISM particles above. 

\subsection{Radiative Processes in the Boson Star Interior}
\label{subsec:radiative}
The resulting emission spectra will depend on the specific radiative processes that occur in the plasma within the boson star, as well as how radiation interacts with it, $i.e.$ the optical thickness. Here, we analyze thermal bremsstrahlung of ISM particles, which is the main contribution to the total radiated energy in the temperature range considered. A discussion of other radiative processes that are subdominant is given in Appendix~\ref{sec:app-e}.

We start by commenting on the optical depth of ISM matter within the boson star. This will be small as long as
\begin{equation}
    2 \sigma_T n R_* \lesssim 1,
\end{equation}
where now $\sigma_T \simeq 10^{-25} \ \rm cm^2$ is the non-relativistic Thomson cross-section, and as before we have conservatively set $n$ to the central ISM particle number density. We find this condition is fulfilled throughout the entire parameter space considered in this work. Therefore, emitted photons rarely scatter against the internal plasma and we expect a bremsstrahlung-type spectrum. In the non-relativistic limit, the emission of final state radiation in $e^- p$ collisions has a specific emissivity given by
    \begin{equation}
      j(E_\gamma)= \frac{16 \pi^2 e^6}{3^{3/2} m_e^2} n^2 \left(\frac{2 m_e}{\pi T}\right)^{1/2}\exp\left(-\frac{E_\gamma}{T}\right),
      \label{eq:bremrate1}
     \end{equation}
where $m_e$ is the electron mass, and $n$ and $T$ respectively are the density and temperature of the charged particles, obtained through Eqs.~\eqref{eq:ISMaccret1} and \eqref{eq:ISMaccret2}. The above expression assumes equal temperature for both charged particle species. However, while protons from the WIM or the HIM phases can be accelerated to substantial energies, electrons are not directly coupled to the bosonic field in this simplified model. Since collisions are rare in this regime, electron thermalization with the boosted ISM baryons is not guaranteed, and therefore may have a different temperature. While in a one-temperature plasma the radiated photon energy comes from the kinetic energy of the scattered electron, when the temperature gap between electrons and protons exceeds $T_{p} \gtrsim (m_n/m_e) \ T_e$, with $m_n/m_e \simeq 1834$, the above emissivity must be corrected to account for the fact that the radiated energy is extracted from the kinetic energy of the protons \cite{Mayer:2006vb,2003A&A...406...31H}. In this limit, the emissivity receives a correction factor of order $(m_e/m_n)^{1/2} \simeq 10^{-2}$ that accounts for the reduction in the mean duration of encounters between electrons and ions. The number of radiated photons $N_\gamma$ per unit volume, time and energy, is given by
     \begin{equation}
       \left(\frac{d^3N_\gamma}{dVdtdE_\gamma}\right)_{\rm brem} \simeq E_{\gamma}^{-1}\left(\frac{m_e}{m_n}\right)^{1/2} j(E_\gamma)~.
     \label{eq:bremrate2}
     \end{equation}
To be conservative in our estimates, we will use Eq.~\eqref{eq:bremrate2} to compute the resulting diffuse photon flux. We remark that radiation in $e^-e^-$ and $pp$ collisions is higher-order in nature and therefore subdominant compared to the above.

\section{Constraints from Diffuse X-Ray and Gamma-Ray Observations}
\label{sec:constraints}

\subsection{Boson Star Luminosity}
\label{subsec:lum}
We begin by computing the luminosity of a boson star as it interacts with the surrounding ISM. The scalings of this quantity with the basic boson star parameters ultimately determine the parameter space where constraints on their fraction can be set. As discussed before, the emissivity is dominated by thermal bremsstrahlung of the ISM particles. Therefore, we express the luminosity in a given photon band as
\begin{equation}
    L_* \simeq \int_{E_\gamma^{\rm min}}^{E_\gamma^{\rm max}} \int_{V_*} E_\gamma \left(\frac{d^3N_\gamma}{dVdtdE_\gamma}\right)_{\rm brem} dV \ dE_\gamma~,
    \label{eq:luminosity}
\end{equation}
where $V_*$ is the boson star volume and we assume that typically $E_{\gamma}^{\rm max} \gg E_{\gamma}^{\rm min}$ for the integrated photon energy. The basic parametrics of the luminosity can be understood in the limit that $T_{\rm ISM} \ll E_\gamma^{\rm max} \ll |V_{\phi}(r)|$ throughout most of the boson star volume. In this case, the exponential tail of the bremsstrahlung spectrum may be neglected and approximate scaling relations are recovered
\begin{equation}
    L_* \sim E_\gamma^{\rm max} R_*^3 \ |V_{\phi}(0)|^{1/2} \propto
    \begin{cases}
      \ m^{-4}M_*^{-1}\Lambda^{-1}  & |\gamma| \lesssim 1  \\
      \\
      \ m^{-4}M_*^{-1}\Lambda^{-1}\gamma^{3/2} & 1 \ll \gamma \lesssim \gamma_{\rm max} \\
    \end{cases}   
    \label{eq:bslum}
\end{equation}

\begin{figure}[h!]
\centering 
\centerline{\includegraphics[width=1.15\textwidth]{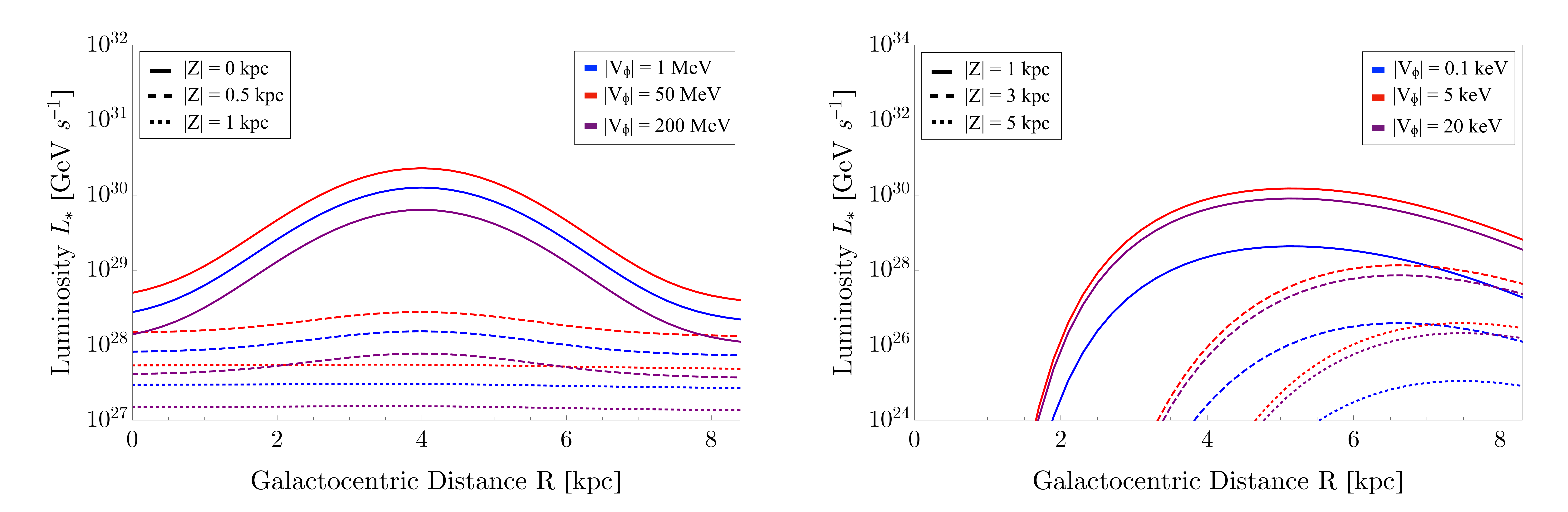}}
\caption{\textbf{Left:} Boson star luminosity in the $(1$, $10) \ \rm MeV$ photon energy band, for boson stars interacting with the WIM, as a function of their galactocentric distance $R$ and for various vertical offsets $Z$ in the Milky Way. We assume a characteristic temperature of $T_{\rm WIM} \simeq 8000 \ \rm K$ for this ISM phase. For concreteness, we have set a boson star mass $M_* = 10^{-7} \ M_{\odot}$, a cutoff scale $\Lambda = 2 \ \rm TeV$, and varied the boson mass $m \gtrsim 10^{-10} \ \rm eV$ to obtain the potential specified for each curve ($cf.$ Eq.~\eqref{eq:nrpotential}). \textbf{Right:} Boson star luminosity in the $(1$, $10) \ \rm keV$ photon energy band, for boson stars interacting with the HIM. We assume a characteristic temperature of $T_{\rm HIM} \simeq 10^6 \ \rm K$ for this ISM phase. In this case, we have set a boson star mass $M_* = 10^{-5} \ M_{\odot}$, a cutoff scale $\Lambda = 2 \ \rm TeV$, and varied the boson mass $m \gtrsim 10^{-13} \ \rm eV$ to obtain the potential specified for each curve.}
\label{fig:lum-ISM}
\end{figure}

The top case corresponds to boson stars with no/weak self-interactions, $cf.$ Eq.~\eqref{eq:gammaweak}, whereas the bottom case corresponds to strong self-interactions, $cf.$ Eq.~\eqref{eq:gammaTF}. The maximum self-interaction parameter for which the above scalings are valid is $\gamma_{\rm max} \propto (m_n/E_{\gamma}^{\rm max})^{1/3} G M_*^{4/3} m^{4/3} \Lambda^{-2/3}$. The luminosity scalings with the basic boson star parameters can be understood as follows: a stronger potential $|V_\phi(0)|$ yields more emission in a given band. However, for a stronger field configuration to be attained, the boson star must be more compact. This implies that, as the potential increases, a peak luminosity is reached when $|V_{\phi}(0)| \gtrsim E_{\gamma}^{\rm max}$, and decreases afterwards, as more compact boson stars enclose less ISM particles that radiate from its interior. This occurs when the self-gravity of the bosons dominates over their quantum pressure, which is reflected in the steep $m^{-4}$ dependence. Conversely, stronger self-interactions lead to more extended boson stars with weaker fields because of the increase in the classical pressure. While $|V_{\phi}(0)| \gtrsim E_{\gamma}^{\rm max}$, this results in a net increase in the luminosity due to the boson star enclosing more ISM particles, which is reflected in the $\propto \gamma^{3/2}$ scaling. However, when $\gamma \gtrsim \gamma_{\rm max}$, the boson star becomes so extended that $|V_{\phi}(0)| \lesssim E_{\gamma}^{\rm max}$ and the exponential tail of the spectrum cannot be neglected. As a result, past this threshold the luminosity steeply decreases. This is illustrated in Figure~\ref{fig:lum-ISM}, where we show the luminosity of a boson star in hard x-ray and soft gamma-ray bands as a function of its Galactic position, for increasing potential values. For concreteness, we have fixed a boson star mass in each case and increased the dark matter mass to obtain the potential specified for each curve.
Based on this, we will see below that our constraints become more stringent for lighter dark matter masses and more massive boson stars, provided the potential sourced $|V_{\phi}(0)| \propto M_*^4 m^4 \gamma^{-3/2}$ is always suitable for the ISM matter to emit x-rays and gamma-rays from within.

The dependence of the luminosity on the Galactic coordinates is directly related to the density distribution of the ISM phases, $cf.$ Figure~\ref{fig:ISMstruc}. In the case of the WIM, the most luminous boson stars lie within the dense annular component that peaks around $R \simeq 4 \ \rm kpc$ and rapidly decays with increasing vertical offset from the galactic midplane. The HIM by contrast is vertically far more extended, and has a much stronger radial variation of its scale height. The brightest boson stars occur at radial distances $R \gtrsim 4 \ \rm kpc$, where the concentration of hot gas is inferred from the supernova rate and hydrostatic equilibrium considerations near the galactic midplane. In general, it can be seen that the luminosity values are tiny compared to typical galactic x-ray sources, indicating that boson stars within the most rarefied ISM phases will be too faint to be directly detected through this effect for most of their parameter space. Below, we analyze the associated flux from both individual boson stars and the full Galactic population within the WIM and the HIM.

\subsection{Diffuse Photon Flux}
\label{subsec:diff}
Diffuse x-ray and gamma-ray data from various surveys has been used in the past to derive constraints on light dark matter decay and annihilation for various well-motivated models \cite{Essig:2013goa,Laha:2020ivk,Cirelli:2020bpc}. In particular, we will follow here a similar analysis to \cite{Essig:2013goa}. The flux spectrum of an individual boson star is
\begin{equation}
   \left(\frac{d^{2}\Phi_{\gamma}}{d\Omega dE_{\gamma}}\right)_* = \frac{1}{4\pi s^2} \int_{V_{*}} \left(\frac{d^3N_\gamma}{dVdtdE_\gamma}\right)_{\rm brem} dV~,
   \label{eq:diff-flux-1}
\end{equation}
where we have abbreviated $\Phi_\gamma \equiv d^{2}N_{\gamma}/dAdt$. In the above integral, $V_{*}$ is the boson star volume and $s$ is its distance to the Earth. With the possible exception of boson stars within the Local Bubble itself, the resulting energy flux in x-rays and gamma-rays is $\lesssim 10^{-16} \ {\rm erg \ cm^{-2} \ s^{-1}}$, which lies below the detection threshold of existing or upcoming instruments. Further details on the detection prospects of boson stars as point sources can be found Appendix~\ref{sec:app-f}. By contrast, the copious amount of boson stars located in the large volumes of WIM and HIM can produce a sizable diffuse flux. We compute this flux by summing the contributions of all boson stars within line of sight (l.o.s), 
\begin{equation}
    \left(\frac{d^{2}\Phi_{\gamma}}{d\Omega dE_{\gamma}}\right)_{\rm tot}=f_{\rm ISM} \ f_* \int_{\rm l.o.s} \left(\frac{\rho_X(s)}{M_*}\right) \left(\frac{d^{2}\Phi_{\gamma}}{d\Omega dE_{\gamma}}\right)_* s^2 ds d\tilde{\Omega}~,
    \label{eq:diff-flux-2}
\end{equation}
where $\rho_X(s)$ is the distribution function of the dark matter, $f_*$ is the fraction of dark matter in boson stars, and we assume for simplicity that all boson stars have approximately the same mass $M_*$. The quantity $f_{\rm ISM}$ is the volume filling factor of the ISM phase under consideration, roughly $f_{\rm WIM} \simeq 0.15$ and $f_{\rm HIM} \simeq 0.5$. The differential solid angle of the field of view is $d\tilde{\Omega} = \sin b \ db \ dl$, where $(l,b)$ are the galactic longitude and latitude, and the domain of integration is specified further below. Note that we only consider the diffuse photon flux produced by Galactic boson stars. The contribution from extra-galactic boson stars is both subdominant and far more uncertain, and is therefore excluded from this analysis. 
In practice, Eq.~\eqref{eq:diff-flux-2} must be numerically integrated for a given dark matter and ISM distribution. The latter is implicitly included in the emissivity given by Eq.~\eqref{eq:bremrate2}, which in turn depends on the ISM temperature and density through Eqs.~\eqref{eq:ISMaccret1} and \eqref{eq:ISMaccret2}. The structure of the ISM is incorporated through Eqs.~\eqref{eq:HIMstruc} and \eqref{eq:WIMstruc} for the HIM and the WIM respectively, see Appendix~\ref{sec:app-a} for details. For the dark matter masses we consider in this work, the de Broglie and Compton wavelengths of the bosons are much shorter than $\sim 10^{-10} \ \rm pc$, therefore we neglect quantum interference effects and assume for simplicity a Navarro-Frenk-White profile for the dark matter halo \cite{Navarro:1995iw,Navarro:1996gj}, with an inner slope $\alpha = 1$, scale radius $r_s = 20 \ \rm kpc$, and scale density $\rho_s = 8 \times 10^6 \ M_{\odot} \ \rm kpc^{-3}$ (see $e.g.$ \cite{Schive:2014hza} for further details on density profiles of lighter dark matter).

\begin{figure}[h!]
\centering 
\centerline{\includegraphics[width=1.20\textwidth]{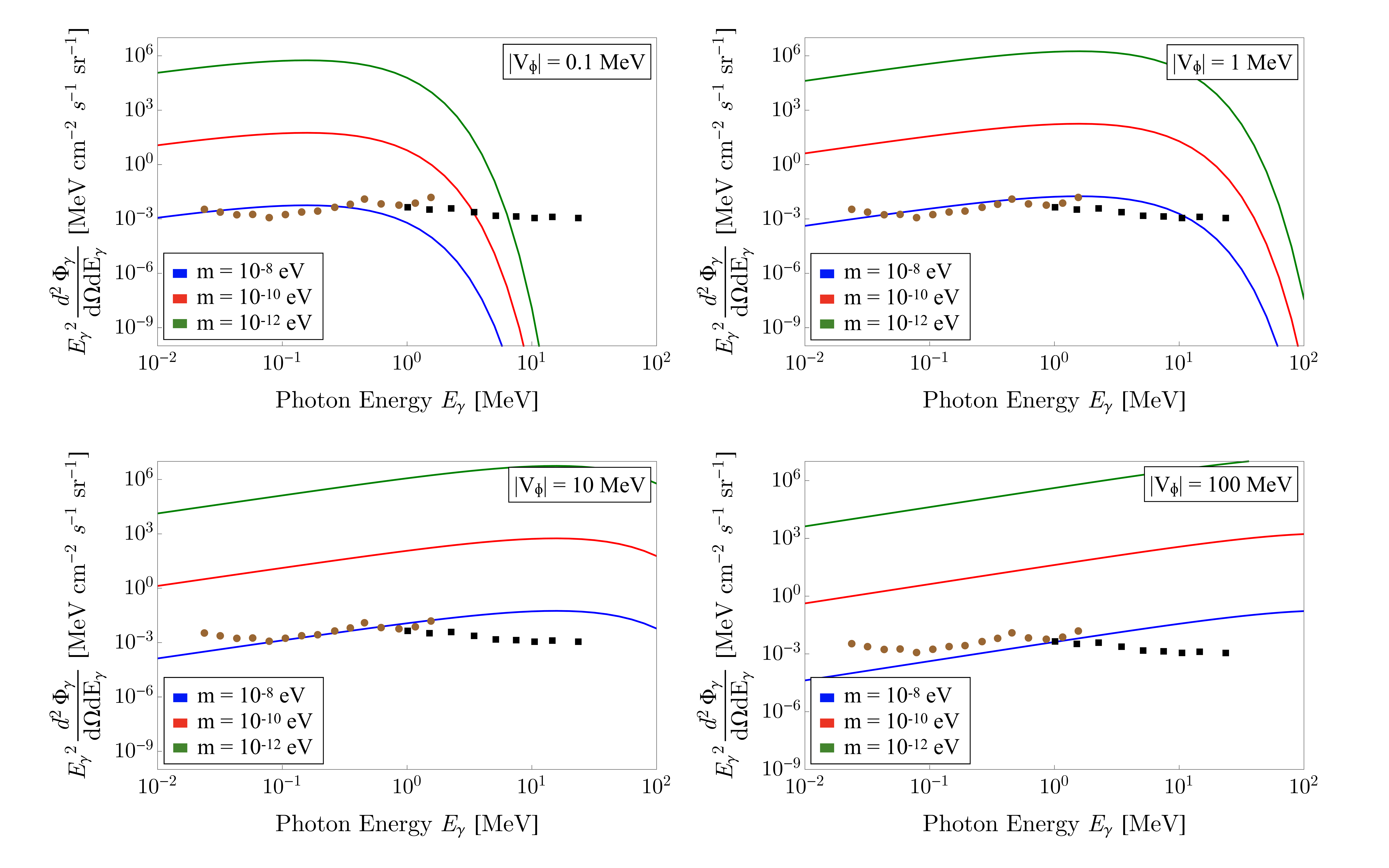}}
\caption{Diffuse flux spectra for boson stars interacting with the WIM, for various boson masses and different potential values as specified in each panel. For these curves, we have fixed a cutoff scale of $\Lambda = 2 \ \rm TeV$ and assumed that boson stars make up all of dark matter, $i.e.$ $f_*=1$. We assume the {WIM} has a volume filling factor of $f_{\rm WIM} \simeq 0.15$, a characteristic temperature $T_{\rm WIM} \simeq 8000 \ \rm K$ and a Galactic distribution given by Eq.~\eqref{eq:WIMstruc}. The data points are diffuse all-sky gamma-ray spectra measured by {INTEGRAL} (\textbf{brown circles}) \cite{Bouchet:2008rp} and {COMPTEL} (\textbf{black squares}) \cite{1998PhDT.........3K}, and are the central value plus $2$x the vertical error bar reported (see text for specific details on each measurement). The spectra were obtained by integrating Eq.~\eqref{eq:diff-flux-2} over the solid angle subtended by {INTEGRAL}'s observation. Thus, to allow proper comparison with the curves, we have appropriately re-scaled {COMPTEL}'s data for these plots by a geometric factor $\int_{\rm INT} n_{\rm WIM}^2(s) \rho_X(s) ds d\tilde{\Omega} \ / \int_{\rm COMP} n_{\rm WIM}^2(s) \rho_X(s) ds d\tilde{\Omega} \sim 0.5$.}
\label{fig:spec-warm-ISM}
\end{figure}

\begin{figure}[h!]
\centering 
\centerline{\includegraphics[width=1.20\textwidth]{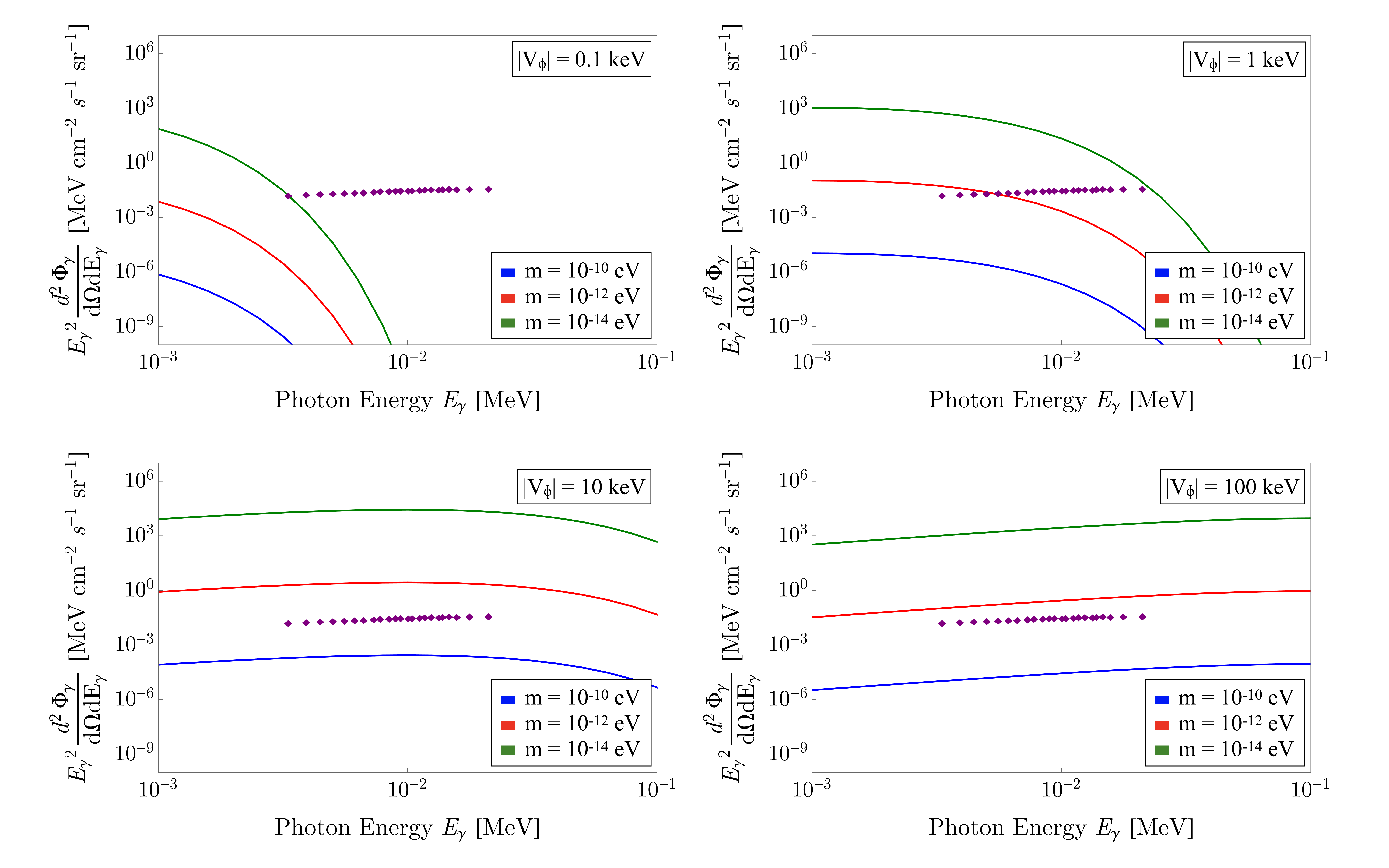}}
\caption{Same as Figure~\ref{fig:spec-warm-ISM}, but for boson stars interacting with the HIM. We assume the {HIM} has a volume filling factor of $f_{\rm HIM} \simeq 0.5$, a characteristic temperature $T_{\rm HIM} \simeq 10^6 \ \rm K$, and a Galactic distribution given by Eq.~\eqref{eq:HIMstruc}. The data points are diffuse x-ray background observations performed by {HEAO-1} (\textbf{purple diamonds}) \cite{Gruber:1999yr}. The spectra were obtained by integrating Eq.~\eqref{eq:diff-flux-2} over the solid angle subtended by this observation.}
\label{fig:spec-hot-ISM}
\end{figure}

For this work, we compare the diffuse photon flux given in Eq.~\eqref{eq:diff-flux-2} to previous observations in hard x-rays and soft gamma-rays performed by {HEAO-1} \cite{Gruber:1999yr}, {INTEGRAL} \cite{Bouchet:2008rp} and {COMPTEL} \cite{1998PhDT.........3K}. 
Specifically, {HEAO-1} has measured the diffuse photon flux spectrum in the $(3$, $70) \ \rm keV$ band in the sky region $b \in (-90$, $-20) \cup (20$, $90) \ \deg$ and $l \in (58$, $109) \cup (238$, $289) \ \deg$; {INTEGRAL} has performed a diffuse photon survey in the $(0.02$, $1) \ \rm MeV$ band in the region $l \in (-15$, $15) \ \deg$ and $b \in (-30$, $30) \ \deg$; and {COMPTEL} has made a similar observation in the band $(1$, $15) \ \rm MeV$ and coordinates $l \in (-20$, $20) \ \deg$ and $b \in (-60$, $60) \ \deg$.
The sharp decay of the {WIM} density with galactic latitude, $cf.$ Eq.~\eqref{eq:WIMstruc}, implies that the diffuse photon data from {HEAO-1} should be compared to the flux produced by boson stars interacting with the {HIM}. By contrast, near the galactic plane, we find that the dominant contribution to the photon flux comes from boson stars situated in the {WIM}, since this phase is more prevalent than its hot counterpart. 
Therefore, we will compare {HEAO-1} data to the flux produced by boson stars residing exclusively in the {HIM}; and {INTEGRAL} and {COMPTEL} data to the flux produced by boson stars residing exclusively in the {WIM}. 
To derive conservative constraints on $f_*$ below, for a fixed cutoff scale $\Lambda$, we will consider the central value plus twice the vertical error bar reported for the {INTEGRAL} and {COMPTEL} surveys. {HEAO-1} data, on the other hand, has negligibly small error bars and places overall weaker constraints. Given the uncertainties in the HIM modeling particularly at high Galactic latitudes, as well as the possibility of x-ray extinction due to dust, the resulting constraints will also be less robust, see Appendix~\ref{sec:app-a} for further details. The data sets used are displayed in Figures~\ref{fig:spec-warm-ISM} and \ref{fig:spec-hot-ISM} along with several sampled diffuse fluxes for comparison. 

Approximate scalings of the diffuse photon spectrum with the basic boson star parameters can also be obtained when assuming a uniform potential throughout the boson star volume that is much greater than the thermal energy of the ISM particles, $i.e.$ $T_{\rm ISM} \ll E_{\gamma} \ll |V_\phi(r)|$ throughout most of the star. As before, these are given separately in the weak and strong self-interaction cases,
\begin{equation}
    \left(\frac{d^{2}\Phi_{\gamma}}{d\Omega dE_{\gamma}}\right)_{\rm tot} \propto \frac{R_*^3 \  |V_\phi(0)|^{1/2}}{M_* E_\gamma} \propto
    \begin{cases}
      \ m^{-4}M_*^{-2}\Lambda^{-1}  & |\gamma| \lesssim 1  \\
      \\
      \ m^{-4}M_*^{-2}\Lambda^{-1} \gamma^{3/2} & 1 \ll \gamma \lesssim \gamma_{\rm max} \\
    \end{cases}   
    \label{eq:diff-flux-3}
\end{equation}
$\gamma_{\rm max}$ was defined below Eq.~\eqref{eq:bslum}. Compared to Eq.~\eqref{eq:bslum}, the extra power in $M_*$ arises from the factor $\rho_X(s)/M_*$ in Eq.~\eqref{eq:diff-flux-2}, which accounts for the number of boson stars within line of sight. The maxima of the energy flux spectra occur at a photon energy $E_\gamma \sim |V_\phi|$, and are more prominent for larger boson stars given a fixed potential value. Figures~\ref{fig:spec-warm-ISM} and \ref{fig:spec-hot-ISM} show how the spectrum transitions to this scaling regime as the potential increases, for boson stars producing gamma-rays in the WIM using the solid angle spanned by INTEGRAL's observation and x-rays in the HIM using the solid angle spanned by HEAO-1's observation, respectively. For concreteness, we have fixed $f_* = 1$ and an effective coupling scale $\Lambda = 2 \ \rm TeV$. These plots also illustrate that, for dark matter masses $m \lesssim 10^{-8} \ \rm eV$, sufficiently heavy boson stars can produce sizable photon fluxes in the ISM, even for effective boson-nucleon couplings smaller than experimental limits, $cf.$ Eq.~\eqref{eq:nongcoupling} and surrounding discussion.

\begin{figure*}[!th]
    \centering
     \includegraphics[width=0.495\linewidth]{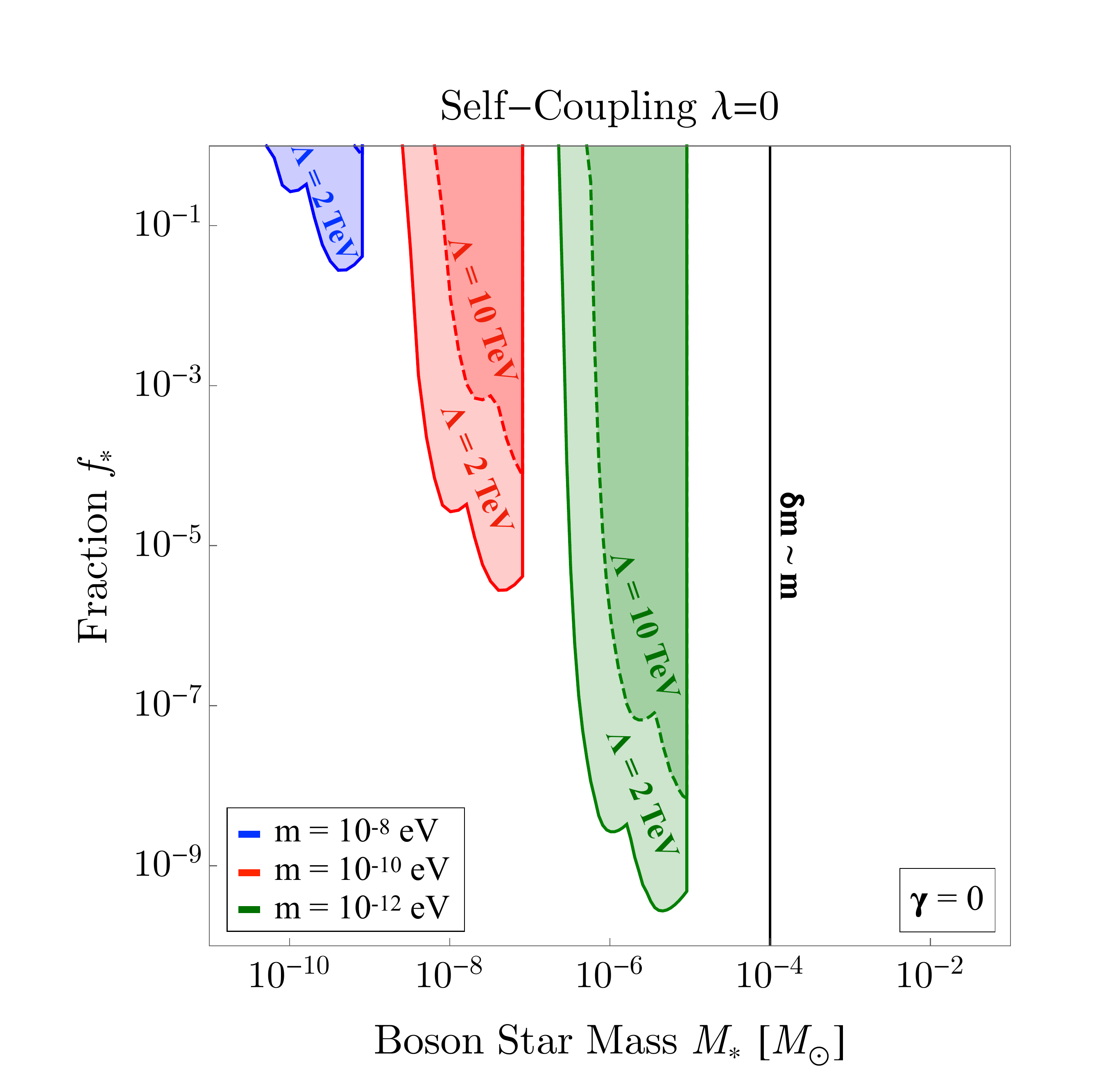}
     \includegraphics[width=0.495\linewidth]{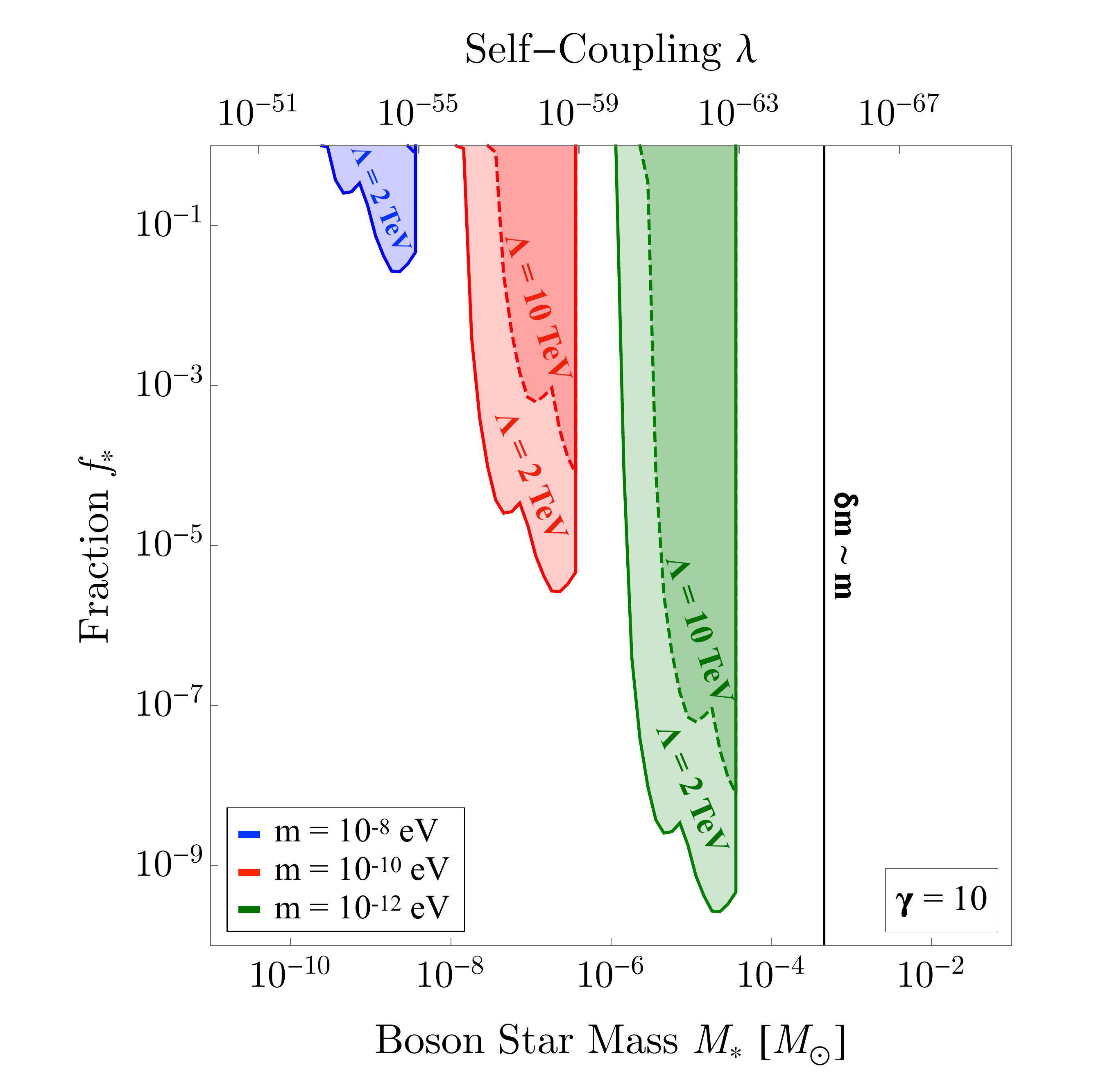}
     \includegraphics[width=0.495\linewidth]{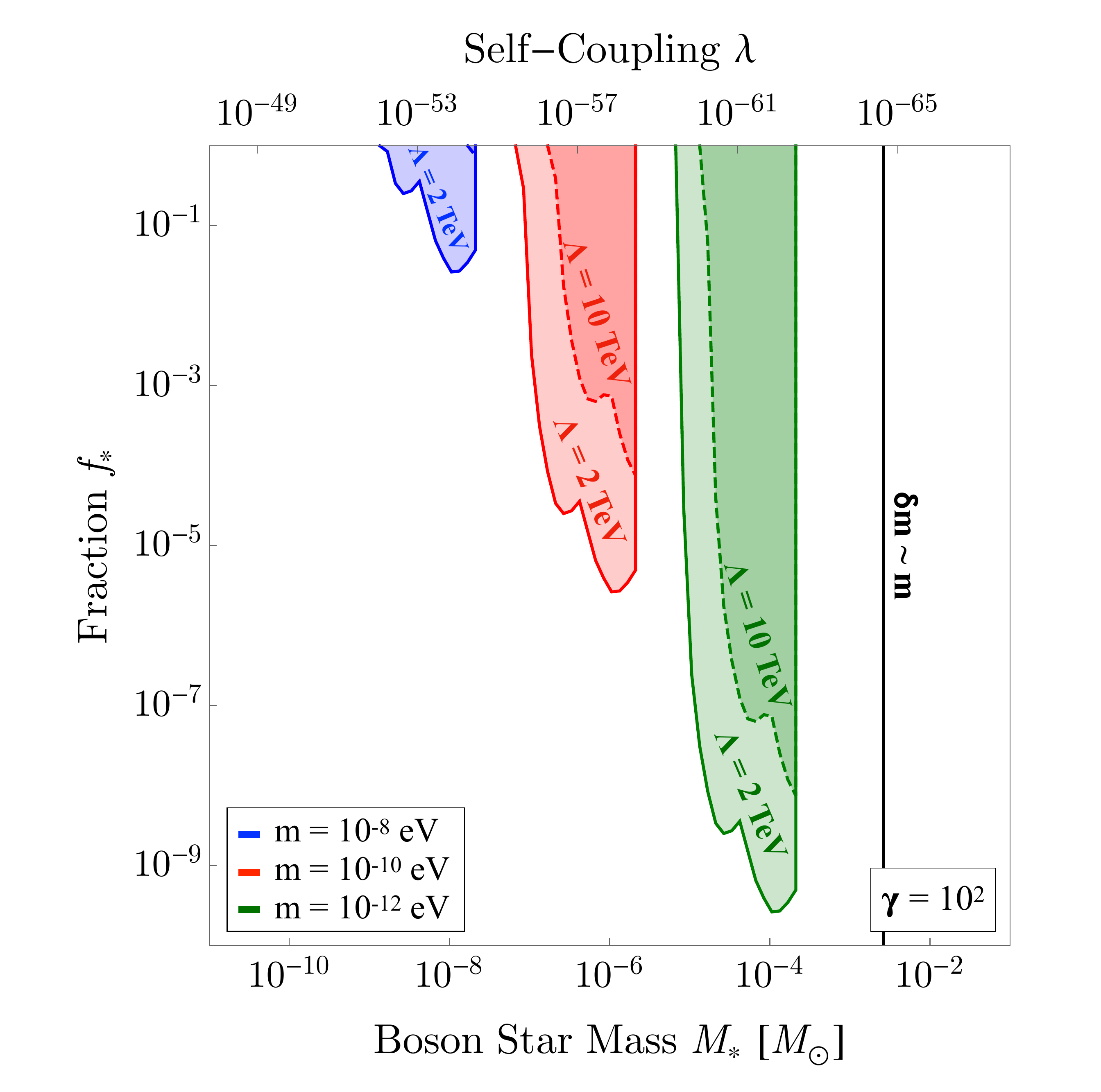}
     \includegraphics[width=0.495\linewidth]{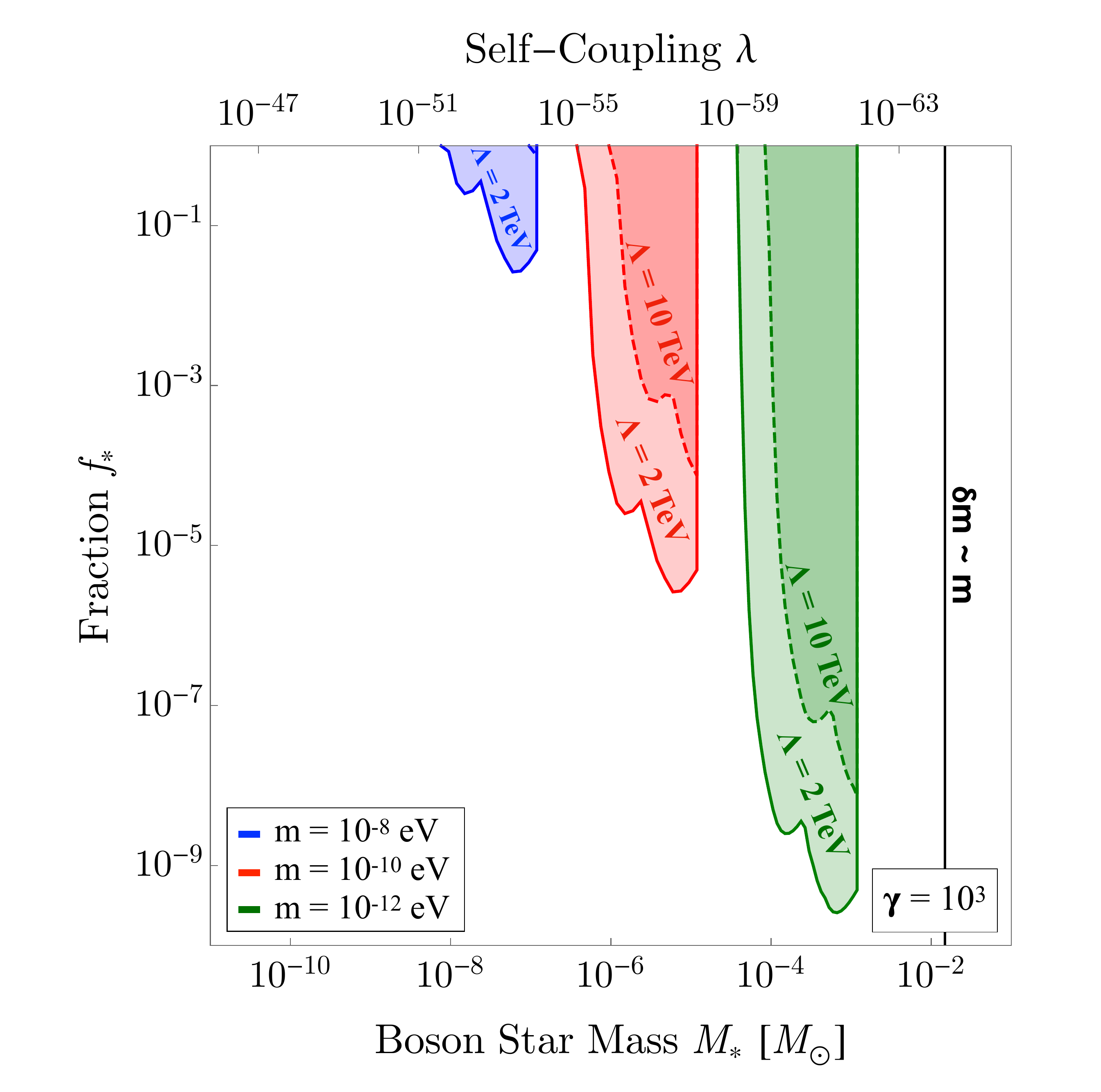}
    \caption{Constraints on the fraction of DM in boson stars, for various dark matter masses, based on the diffuse gamma-ray flux they would produce in the WIM compared to observations performed by {INTEGRAL} and {COMPTEL}. The upper horizontal axes denote the repulsive self-interaction strength $\lambda$, which is set by the self-interaction parameter $\gamma$ in each panel, $cf.$ Eqs.~\eqref{eq:gammaweak} and \eqref{eq:gammaTF}. The vertical line labeled `$\delta m \sim m$' (\textbf{black}) indicates the threshold for which the boson mass correction due to the interior baryonic density becomes significant. In every case, the bounds are terminated to the right at a boson star mass for which the potential sourced becomes relativistic, $i.e.$ $|\phi| \gg \Lambda$ within the boson stars.}
    \label{fig:fractionbounds1}
\end{figure*}

\subsection{Limits from Diffuse X-Ray and Gamma-Ray Data}
\label{subsec:limits}
\begin{figure*}[!th]
    \centering
     \includegraphics[width=0.495\linewidth]{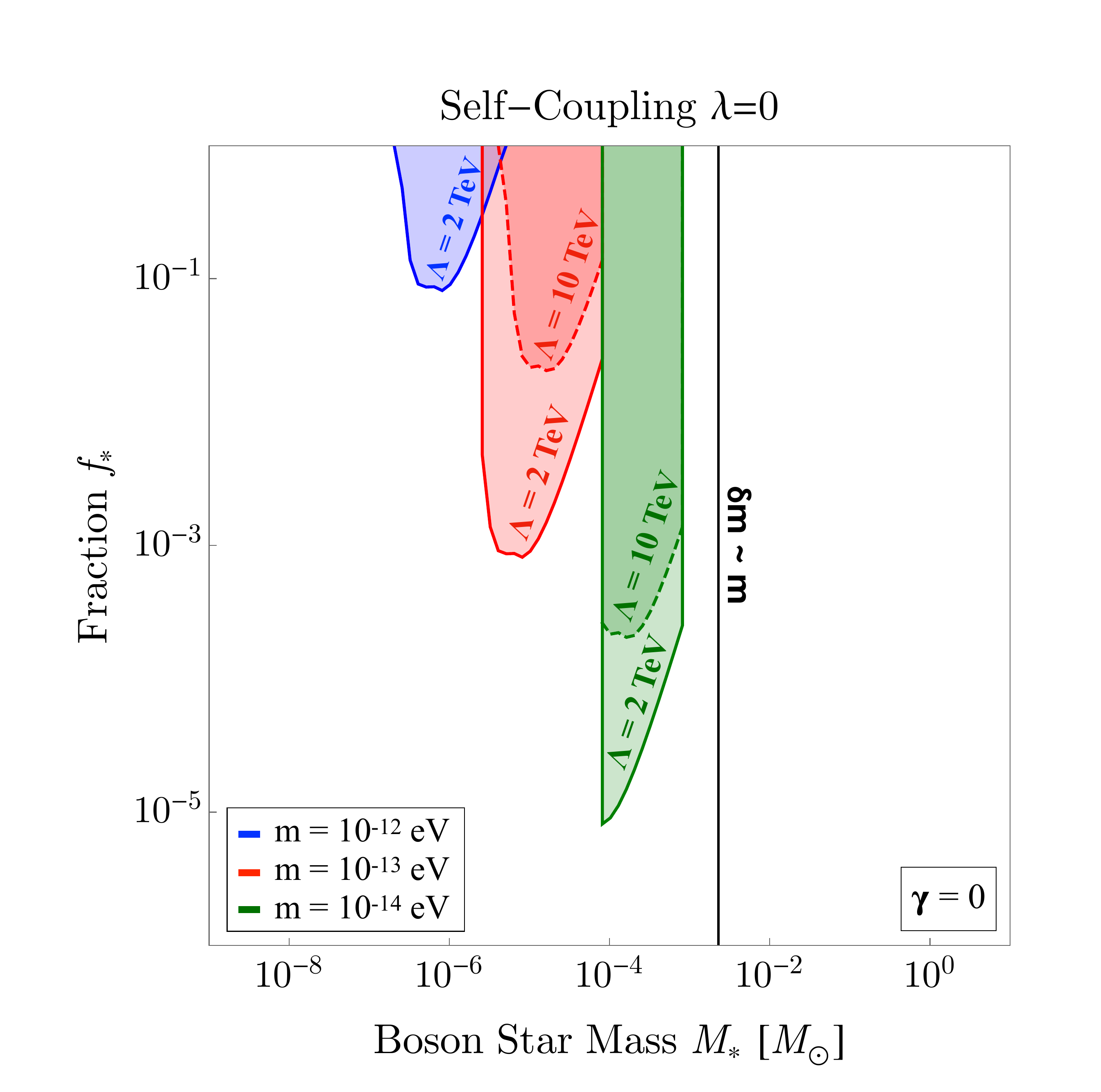}
     \includegraphics[width=0.495\linewidth]{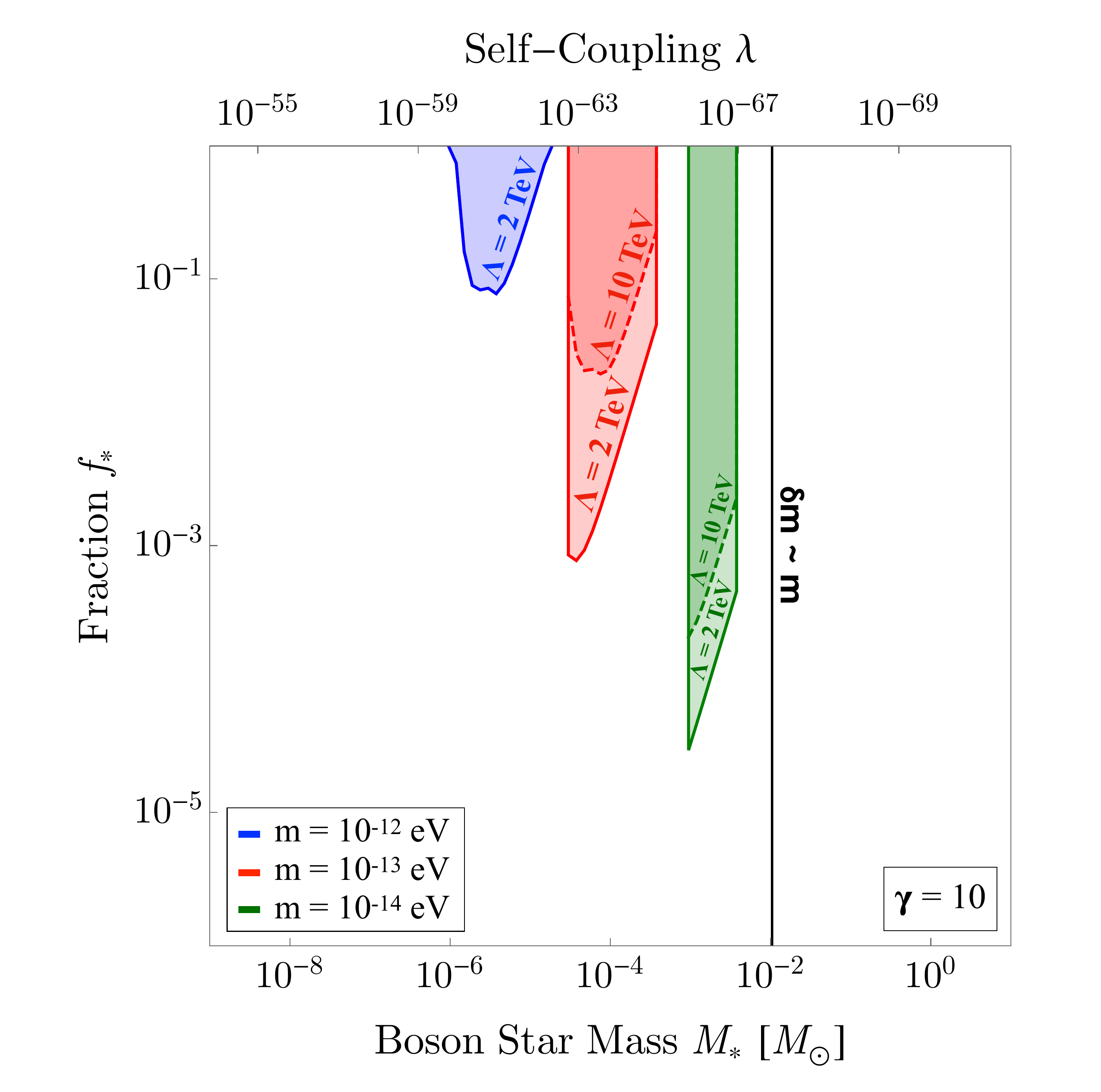}
     \includegraphics[width=0.495\linewidth]{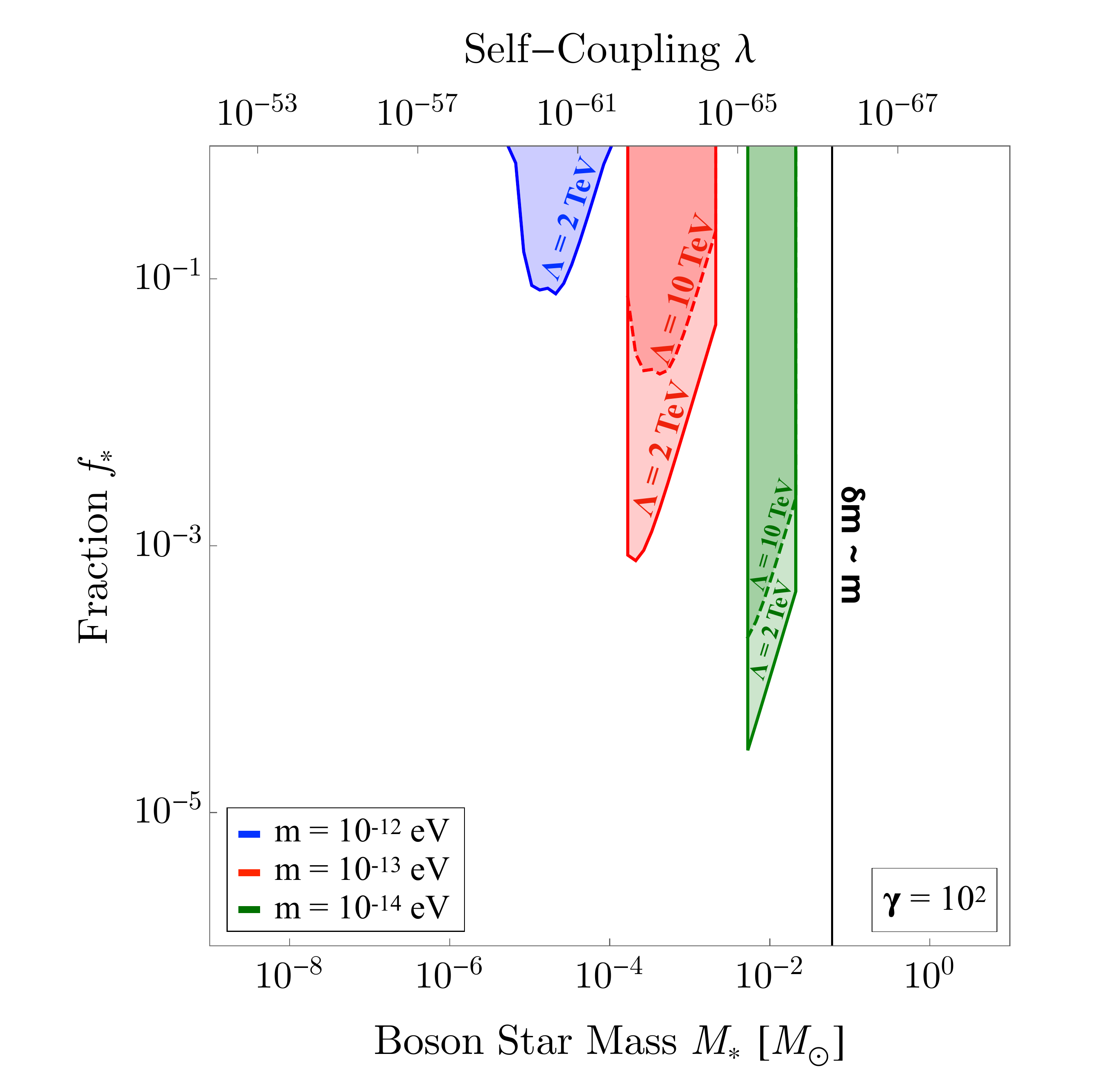}
     \includegraphics[width=0.495\linewidth]{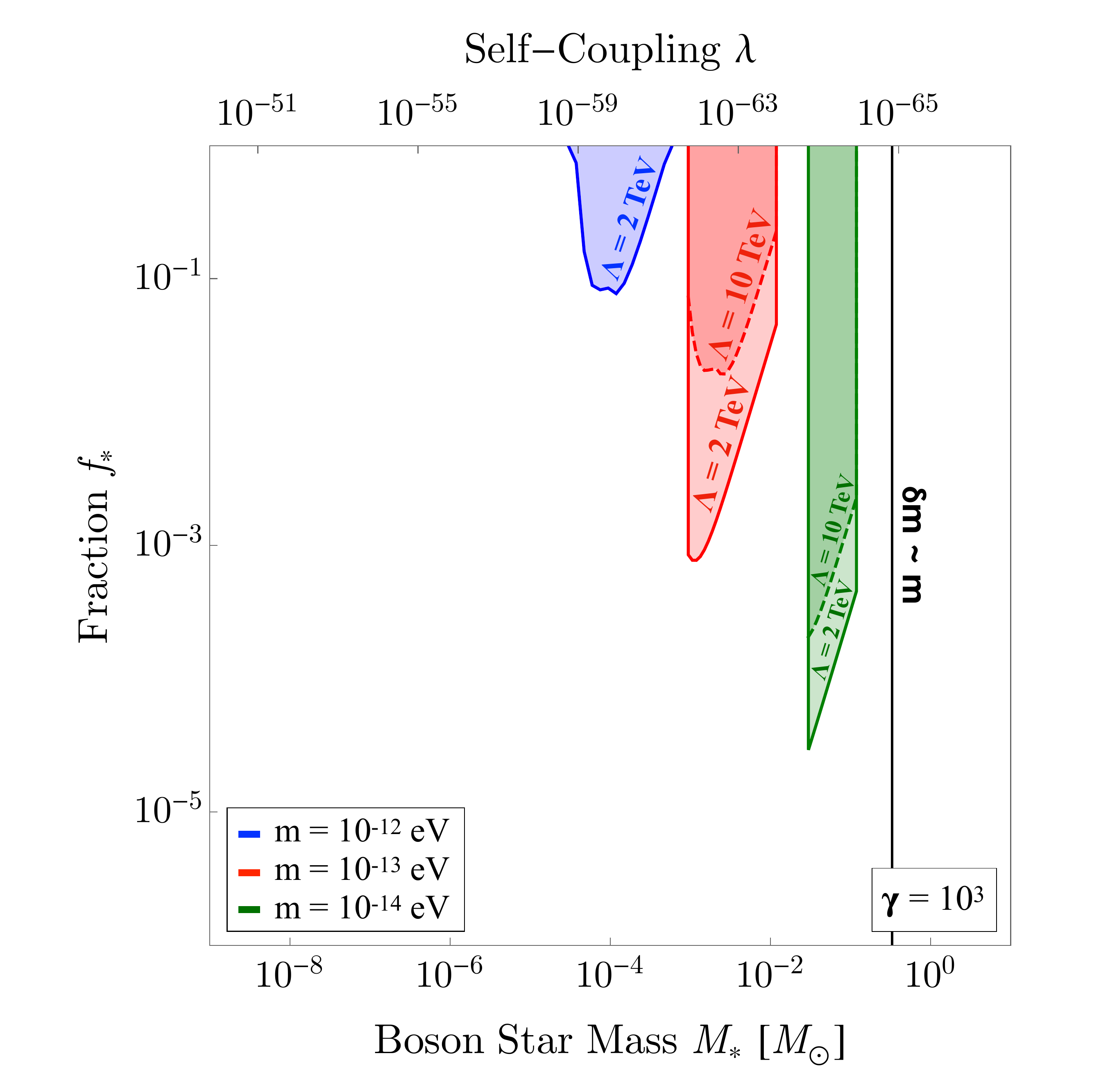}
    \caption{Same as Figure~\ref{fig:fractionbounds1}, but the constraints are derived based on the diffuse photon flux boson stars would produce in the HIM compared to observations performed by HEAO-1. The bounds are terminated to the right at a boson star mass for which the potential sourced becomes relativistic, $i.e.$ $|\phi| \gg \Lambda$ within the boson stars, while to the left they are terminated when boson star radii exceed the mean separation in the halo, $i.e.$ $R_* \gtrsim (\rho_{\odot}/M_*)^{-1/3}$ where $\rho_{\odot} \simeq 10^7 \ M_{\odot} \ \rm kpc^{-3}$ is the local dark matter density.}
    \label{fig:fractionbounds2}
\end{figure*}

Figures~\ref{fig:fractionbounds1} and \ref{fig:fractionbounds2} respectively show the inferred limits on the fraction $f_*$ of dark matter in boson stars, based on the diffuse x-ray/gamma-ray flux they would produce in the WIM and the HIM, for various dark matter masses and self-interaction strengths. For robustness, the constraints are set by computing the minimum $f_*$ value for which the dark matter signal, $i.e.$ the diffuse flux, is above the observed flux plus twice the associated vertical error bar. In principle, stronger constraints could be derived with an appropriate modeling of backgrounds and the use of more sophisticated statistical methods \cite{Essig:2013goa}. To show our results, we have chosen fiducial cutoff scales of $\Lambda = 2 \ \rm TeV$ and $\Lambda = 10 \ \rm TeV$ for the effective operator given by Eq.~\eqref{eq:nongcoupling}, which lie below robust experimental limits based on two-scalar-mediated nucleon-nucleon scattering. 

For a given dark matter mass, the constraints in the low stellar mass end are determined by requiring the boson stars source a potential suitable for the emission of x-rays and gamma-rays. In particular, for the limits derived from x-ray emission in the HIM, we also require the boson star size not to exceed their average separation in the halo. At the high mass end, the field sourced by boson stars becomes so intense that the non-relativistic treatment we are assuming breaks down, $i.e.$ $|\phi| \ll \Lambda$ or equivalently  $|V_{\phi}(0)| \gg m_n$, and so the constraints are terminated past that point. We remark that, for decreasing dark matter mass, the constraints are shifted towards the heavier boson star masses since the potential they source must remain fixed to values $|V_{\phi}(0)| \propto M_*^4 m^4 \gtrsim \rm keV$ for x-ray emission and $|V_{\phi}(0)| \propto M_*^4 m^4 \gtrsim \rm MeV$ for gamma-ray emission. At the same time, the limits become more stringent because the boson star radii steeply increase with decreasing dark matter mass since the quantum pressure becomes more dominant while the self-gravity decreases, implying that more ISM matter that radiates is enclosed in these objects. In a similar manner, stronger repulsive self-interactions also shift the constraints to larger boson star mass values. This is because, for boson stars to produce a sizable x-ray or gamma-ray signal, an increasing classical pressure due to self-interactions must be balanced with a stronger self-gravity of the bosonic particles.

For sufficiently light dark matter, however, the mass correction induced by the baryonic density eventually becomes significant, $i.e.$ $\delta m^2 \simeq m^2$ in Eq.~\eqref{eq:deltam}. In other words, back-action effects of the ISM itself on the bosonic field configuration are no longer negligible, potentially altering the flow of ISM particles through it and the associated the radiative processes. In the limit that $T_{\rm ISM} \ll |V_{\phi}(r)|$, this threshold is estimated utilizing the unbound particle density given in Eq.~\eqref{eq:ISMaccret1} (as we note in Appendix~\ref{sec:app-d}, bound ISM particles can introduce at most an order 1-10 correction factor to this estimate). When computing this mass correction, we conservatively assume a cutoff scale $\Lambda = 2 \ \rm TeV$ and a background ISM density of order $0.1 \ \rm cm^{-3}$ for the WIM and $10^{-3} \ \rm cm^{-3}$ for the HIM, although the precise value will depend on the Galactic position of the boson star. This estimate indicates that back-action becomes significant for dark matter masses $m \lesssim 10^{-13} \ \rm eV$ for boson stars that can produce gamma-rays in the WIM and $m \lesssim 10^{-15} \ \rm eV$ for boson stars that can produce x-rays in the HIM.

Finally, although we have not considered boson star parameters for which the non-relativistic treatment breaks down, we remark that diffuse photon constraints could potentially be extended into this region of parameter space. However, the analysis would be significantly more complicated, as other high energy processes and the dynamics of accretion in a relativistic regime would need to be accounted for. In principle, observations of diffuse high energy gamma-rays with $e.g.$ Fermi-LAT could place the most stringent constraints in this case. 

\section{Summary and Conclusions}
\label{sec:concl}
In this work, we have explored a novel indirect search method for boson stars made of light dark matter that has weak effective couplings to the Standard Model nucleon. We summarize our results as follows:
\begin{itemize}
    \item Using well tested semi-analytical solutions for Newtonian boson stars, we have shown how small effective couplings of light scalar dark matter to the Standard Model nucleon can lead to an attractive potential between baryons and the boson star. This potential can be sizable, even for effective couplings that are smaller than current experimental limits. While various signatures of this effect could be analyzed, we have focused here on the phenomenology of ISM baryons intersecting the boson star and gaining kinetic energy while in transit. At the same time, due to the low ISM density this signature is robust to back-action effects of the ambient baryons on the bosonic field configuration.
    
    \item For small boson stars located in the ionized phases of the ISM, the resulting baryon density and temperature distribution can be estimated by modeling the particle flow as approximately collisionless. We find these quantities to be increased within the boson star volume, compared to their boundary ISM values far from the boson star. This in turn implies that radiative processes like thermal bremsstrahlung proceed at an enhanced rate inside the star, and are capable of emitting photons of higher energy compared to the background ISM.
    
    \item This effect, however, would not make the majority of boson stars in the ionized ISM luminous enough to be detected as point sources. Instead, we have found that the diffuse photon flux produced by all the boson stars within the large volumes of warm ionized and hot ISM can be significant. We have computed this diffuse flux in hard x-rays and soft-gamma rays, accounting for both the ISM and dark matter distributions in the Milky Way. 
    
    \item Comparing the resulting flux to existing diffuse x-ray and gamma-ray observations carried out by HEAO-1, INTEGRAL and COMPTEL, we have set strong constraints on the fraction of weakly-coupled bosonic dark matter that can be in boson stars. Specifically, for dark matter masses $(10^{-14}$, $10^{-8}) \ \rm eV$, boson star masses $(10^{-10}$, $10^{-1}) \ M_{\odot}$ and repulsive self-interaction strengths up to $\lesssim 10^{-52}$, the limits we have found can be as stringent as $f_* \lesssim 10^{-9}$ depending on the effective coupling between the dark matter and the nucleon field.
\end{itemize}

To our knowledge, this is the first indirect detection study proposed for boson stars based on diffuse non-transient x-ray and gamma-ray signals, and has relevant implications for the detection of light dark matter with tiny couplings to the Standard Model. Several collaborations have searched for signals of light scalar dark matter in the form of time oscillations or environmental dependence of various physical constants, and have in turn placed strong limits on linear and quadratic dark matter couplings to the Standard Model. Moreover, numerous proposals have the potential to improve upon this in the near future. These experiments, however, depend on the precise configuration of the light dark matter field in their vicinity. If self-interactions, gravitational instabilities, or some other mechanism has caused a sizable amount of dark matter to form compact structures that rarely intersect with the Earth, like boson stars, then such constraints may be weakened or even not be applicable. Here, we have shown for the first time how constraints can be set on the amount of dark matter within such objects, using existing diffuse x-ray and gamma-ray observations and assuming quadratic couplings that are weaker than the most robust experimental limits. At the same time, due to the low density of the ISM, this method has the unique advantage that back-action from the ambient baryonic density on the scalar field configuration is negligible for most of the parameter space, an effect that could negatively impact on the sensitivity of future terrestrial experiments.

Our present work could be extended in a number of ways. For simplicity, we have considered only one specific coupling to the Standard Model nucleon. Effective couplings to leptons, photons and massive gauge bosons can be the subject of future investigation. On the other hand, while we have focused so far on x-ray and gamma-ray signals, it will be interesting to analyze emission in longer wavelengths, which could potentially place more stringent constraints on the fraction and couplings of these objects. This will require a more detailed analysis of backgrounds and how the emitted radiation propagates through the ISM. In addition, point source searches with next-generation instruments may pose an alternative way to find or constrain boson stars. Finally, our work may find applications in other macroscopic dark matter models such as Q-balls, topological defects, vector dark matter structures and fermionic composite states. 

\section*{Acknowledgements}
We would like to thank Rebecca Leane and Aaron Vincent for valuable comments and discussions. The work of JA and JB is supported by the Natural Sciences and Engineering Research Council of Canada (NSERC). Research at Perimeter Institute is supported in part by the Government of Canada through the Department of Innovation, Science and Economic Development Canada and by the Province of Ontario through the Ministry of Colleges and Universities. JA thanks Perimeter Institute for hospitality while portions of this work were completed. 

\appendix

\section{Modeling of the ISM Structure}
\label{sec:app-a}
\begin{figure}[h!]
\centering 
\centerline{\includegraphics[width=1.15\textwidth]{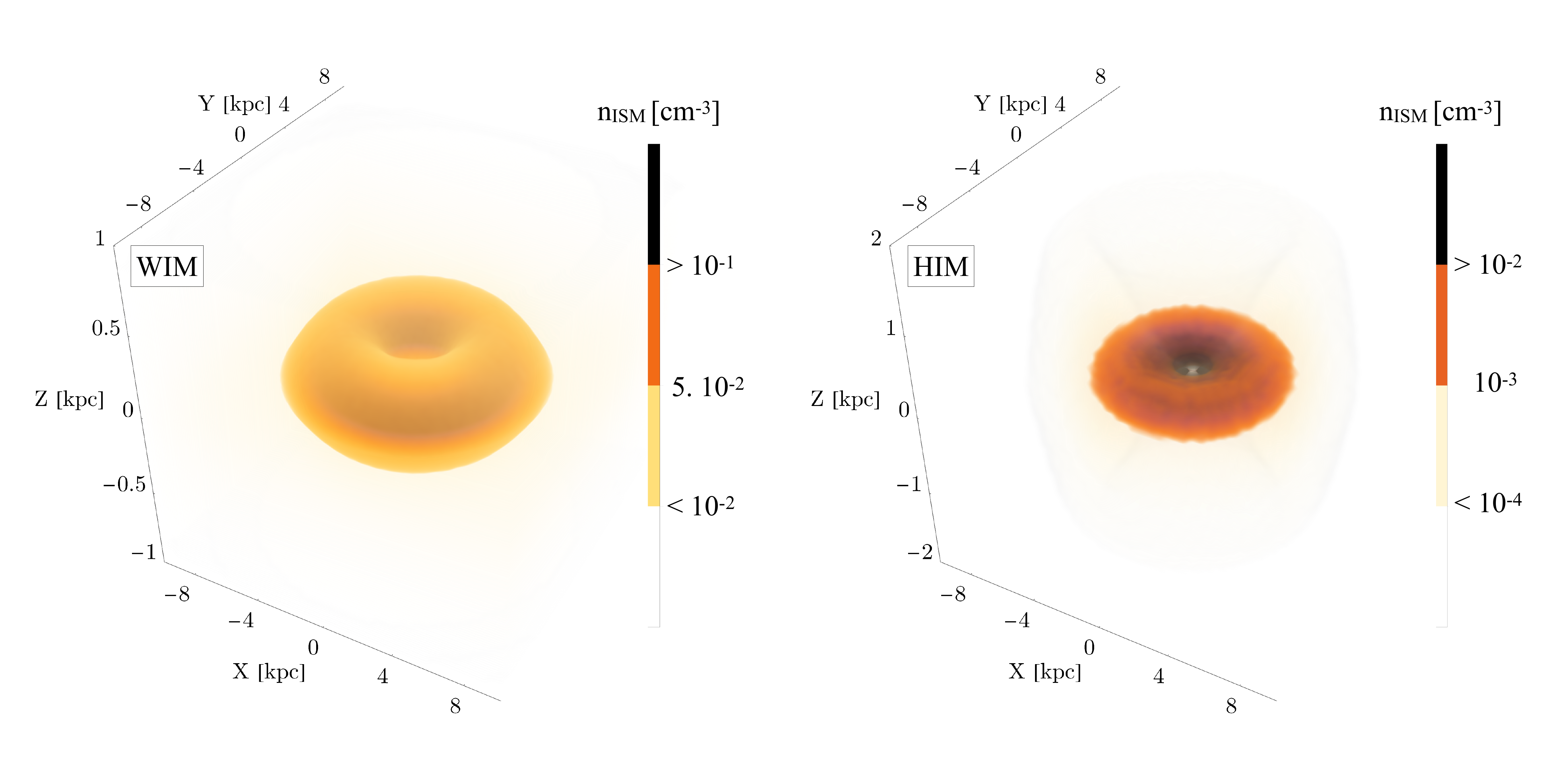}}
\caption{\textbf{Left:} Space-averaged density of the warm ionized interstellar medium (WIM) within the inner $R \lesssim 8 \ \rm kpc$ of the Milky Way, as modeled by \cite{1998ApJ...497..759F}. \textbf{Right:} Same as left panel, but for the hot interstellar medium (HIM) (note the increased vertical scale compared to the left panel).}
\label{fig:ISMstruc}
\end{figure}

In this Appendix, we provide further detail on the ISM phases of relevance for this work. These are the warm ionized ISM (WIM) and hot ISM (HIM) phases, which occupy a significant fraction of the total ISM volume. Their low density implies that Eq.~\eqref{eq:mfp-1} is satisfied for the boson star radii considered, and back-action effects on the boson star field can be neglected for most of their parameters, unlike the colder and denser ISM phases (see below). Both the number density and temperature of these phases are required as boundary conditions for Eqs.~\eqref{eq:ISMaccret1} and \eqref{eq:ISMaccret2}. On the other hand, the spatial distribution of the WIM and the HIM in the Milky Way is needed to compute the diffuse flux produced by boson stars along a certain line of sight, $cf.$ Eq.~\eqref{eq:diff-flux-2}. We detail on their modeling below:
\begin{enumerate}
    \item \textit{Hot Interstellar Medium} (HIM): This ISM phase is formed by hot ejected material from supernovae and other high-energy astrophysical phenomena. Due to its high buoyancy, it is predominant at high galactic latitudes in the form of fountains and bubbles. This phase has a characteristic temperature of order $T_{\rm HIM} \simeq 10^6 \ \rm K$, and a low number density of order $n_{\rm HIM} \simeq 10^{-3} \ \rm cm^{-3}$. The ISM volume that this phase represents is approximately $f_{\rm HIM} \simeq 0.5$. For this work, we will use the Galactic HIM distribution given in \cite{1998ApJ...497..759F}, which assumes a constant HIM temperature of $T_{\rm HIM} \simeq 10^6 \ \rm K$, and the averaged density distribution
    \begin{align}
        \langle n_{\rm HIM}(R,Z) \rangle & \simeq 4.8 \times 10^{-4} \ {\rm cm^{-3}} \left(\frac{R}{R_\odot}\right)^{-1.65} \exp\left(-\frac{|Z|}{H_h(R)}\right) \times \label{eq:HIMstruc} \\ & \left(0.12 \exp\left(-\frac{R-R_\odot}{4.9 \ \rm kpc}\right) + 0.88 \exp\left(-\frac{(R - 4.5 \ {\rm kpc})^2 - (R_\odot - 4.5 \ {\rm kpc})^2}{(2.9 \ \rm kpc)^2}\right) \right)~, \nonumber
    \end{align}
    \begin{equation}
        H_h(R) \simeq 1.5 \ {\rm kpc} \left(\frac{R}{R_\odot}\right)^{1.65}~,
    \end{equation}
    where $R$ is the galactocentric distance ($R_\odot \simeq 8.4 \ \rm kpc$), and $Z$ is the vertical distance to the galactic midplane. The spatial distribution of this phase, over $\sim$kpc scales, is plotted in Figure~\ref{fig:ISMstruc}. We point out that the Solar System is at present transiting this phase of hot rarefied gas, the so-called Local Bubble. In particular, the prevalence of the HIM at high galactic latitudes are relevant for setting constraints on boson star abundance using HEAO-1 observations in the hard x-ray band. At such high galactic latitudes, the hot gas column density is of order $\lesssim 10^{18} \ \rm cm^{-2}$, so we expect x-ray extinction due to dust to be mostly negligible \cite{1995A&A...293..889P}. 

    \item \textit{Warm Ionized Interstellar Medium} (WIM): This ISM phase is mostly composed of diffuse photoionized hydrogen gas, and contains a sizable fraction of the HII in the galaxy. The temperature of this phase is of order $T_{\rm WIM} \simeq 8000 \ \rm K$, with a significantly higher number density compared to the HIM, of approximately $n_{\rm WIM} \simeq 0.05 \ \rm cm^{-3}$. This phase represents approximately a $f_{\rm WIM} \simeq 0.15$ fraction of the total ISM volume. As in the previous case, we use the modeling of \cite{1998ApJ...497..759F}, which assumes a characteristic temperature $T_{\rm WIM} \simeq 8000 \ \rm K$ and an average density distribution
    \begin{align}
        \langle n_{\rm WIM}(R,Z) & \rangle \simeq 0.0237 \ {\rm cm^{-3}} \exp\left(-\frac{|Z|}{1 \ \rm kpc}\right) \exp\left(-\frac{R^2-R_\odot^2}{(37 \ \rm kpc)^2}\right) \label{eq:WIMstruc} \\ & + 0.0013 \ {\rm cm^{-3}} \exp\left(-\frac{|Z|}{0.15 \ \rm kpc}\right) \exp\left(-\frac{(R-4\ {\rm kpc})^2-(R_\odot-4 \ {\rm kpc})^2}{(2 \ \rm kpc)^2}\right). \nonumber
    \end{align}
    This distribution is also shown in Figure~\ref{fig:ISMstruc}. The diffuse photon contribution of boson stars residing in the WIM is dominant compared to the HIM near the galactic midplane, and we use diffuse gamma-ray measurements from INTEGRAL and COMPTEL to constrain the boson star abundance in this region.
\end{enumerate}

The ISM boundary values listed for the hottest phases above, combined with the boson star potential, $cf.$ Eq.~\eqref{eq:nrpotential}, are sufficient to determine the unbound ISM particle number density and temperature within the star. We highlight that, although we assume here a uniform temperature for these phases, our results are insensitive to the precise value of this quantity since appreciable diffuse fluxes are obtained in the limit that $|V_\phi(r)| \gg T_{\rm ISM}$. Most of the uncertainty of our results, within this global ISM modeling, comes from the supernova rate which is used to determine the density of hot ionized gas \cite{1998ApJ...497..759F}. Even so, we expect the associated uncertainty in density to be at most within an order unity factor, which does not significantly impact our results. The use of more accurate ISM models is left for future exploration.

For completeness, we briefly comment on the remaining neutral ISM phases. Ordered from warmest to coolest, these are: \textit{warm neutral medium} (WNM) with $T_{\rm WNM} \simeq 6000 \ \rm K$, $n_{\rm WNM} \simeq 0.5 \ \rm cm^{-3}$ and $f_{\rm WNM}\simeq 0.3$; \textit{cold neutral medium} (CNM) with $T_{\rm CNM} \simeq 100 \ \rm K$, $n_{\rm CNM} \simeq 50 \ \rm cm^{-3}$ and $f_{\rm CNM}\simeq 0.05$; and \textit{molecular clouds} (MC) with $T_{\rm MC} \simeq 20 \ \rm K$, $n_{\rm MC} \simeq 10^3 \ \rm cm^{-3}$ and $f_{\rm MC} \ll 0.01$. Boson star signatures in these denser media have been left for future work, since it would require a careful analysis of accretion in a hydrodynamical regime and in some cases also considering the ISM back-action on the bosonic field. 

\section{ISM Particle Scattering against Dark Matter Bosons}
\label{sec:app-b}
We briefly comment on the possibility of transiting ISM particles scattering against the bosons inside the boson star. As discussed in the main text, boson stars are represented by a classical scalar field $\phi$, overlapped with quantum fluctuations $\delta\phi$ which stand for the collective modes of the condensate. Below the critical temperature for condensation, the effective boson-nucleon coupling given by Eq.~\eqref{eq:nongcoupling} introduces interaction terms with the quantum fluctuations of the form $(m_n/\Lambda^2) \sum_{j=1}^{2} \left(\phi_{j} \bar{n} n \delta\phi_{j} + \bar{n} n \delta\phi_{j}^{2}\right)$, where the index $j$ runs over the number of degrees of freedom. In the case of a complex scalar field, the fluctuation fields correspond to a massive and a gapless mode, whereas in the real scalar case only a single massive mode is present. Using basic parametric arguments, the cross-section for a proton to scatter off these modes in the limit $m \ll m_n$ should scale as,
\begin{equation}
    \sigma_{n\delta\phi} \simeq \frac{m_n^2}{\Lambda^4} \simeq 10^{-44} \ {\rm cm^2} \left(\frac{\Lambda}{10 \ \rm TeV}\right)^{-4}.
\end{equation}
By computing the corresponding mean free path associated with this cross-section, we have verified that baryons transiting boson stars considered in this work infrequently excite collective modes. 

\section{Density and Temperature of Unbound ISM Particles}
\label{sec:app-c}
In this Appendix, we show how Eqs.~\eqref{eq:ISMaccret1} and \eqref{eq:ISMaccret2} are obtained. As detailed in the main text, we assume far from the boson star a Maxwell-Boltzmann distribution $f(E) = n_{\rm ISM} (m/2\pi T_{\rm ISM})^{3/2} \exp\left(-m v^2/2 T_{\rm ISM}\right)$ for the ISM particles. In terms of the dimensionless variable $x = |V_\phi(r)|/T_{\rm ISM}$, Eqs.~\eqref{eq:ISMgen1} and \eqref{eq:ISMgen2} can be integrated by parts to yield

\begin{equation}
    n_{E>0}(r) \simeq n_{\rm ISM} \ e^{x} \left[x^{1/2} e^{-x}+\frac{\sqrt{\pi}}{2}\left(1-{\rm erf}\left(x^{1/2}\right)\right)\right] \sim n_{\rm ISM} \ x^{1/2}
\end{equation}

\begin{equation}
    T_{E>0}(r) \simeq T_{\rm ISM} \ \frac{e^{x} \left[ \left(x^{3/2}+\frac{3}{2}x^{1/2}\right) e^{-x}+\frac{3\sqrt{\pi}}{4}\left(1-{\rm erf}\left(x^{1/2}\right)\right)\right]}{e^{x} \left[x^{1/2} e^{-x}+\frac{\sqrt{\pi}}{2}\left(1-{\rm erf}\left(x^{1/2}\right)\right)\right]} \sim T_{\rm ISM} \ x
\end{equation}

The rightmost expressions are the asymptotic values in the limit $x \gg 1$, which implies a strong attractive potential compared to the thermal energy of the ISM particles far from the boson star. These are obtained using the asymptotic expansion of the error function ${\rm erf}(z \gg 1) \simeq 1 - z^{-1} \exp\left(-z^2\right) / \sqrt{\pi}$, and retaining only the highest order terms in $x$.

\section{Capture of ISM Particles through Inelastic Processes}
\label{sec:app-d}
Eqs.~\eqref{eq:ISMaccret1} and \eqref{eq:ISMaccret2} correspond to the number density and temperature of ISM particles that are unbound to the boson star. Although rare in this regime, inelastic collisions will cause a fraction of the ISM particles to become bound over time. As a result, the density of ISM particles inside the boson star will steadily increase, potentially violating the long mean free path condition set by Eq.~\eqref{eq:mfp-1} as well as producing non-negligible back-action effects on the scalar field configuration. Here we show that, although a fraction of the baryons are captured, the density of bound particles remains low on $\sim \rm Gyr$ timescales for the boson star parameters considered. 

Baryons in the WIM and the HIM respectively have thermal energies of order $\rm eV$ and $\rm 100 \ \rm eV$ far from a boson star. To become bound to the boson star, they must at least lose this energy when passing through. As discussed in the main text, the dominant energy loss process is thermal bremsstrahlung. The cross-section for bremsstrahlung has a complicated dependence on the proton energy in the limit that the proton moves much faster than the electron \cite{2003A&A...406...31H}. However, we estimate a maximum cross-section for emitting photons of energies $E_\gamma^{\rm min} \gtrsim {\rm eV}$ ($E_\gamma^{\rm min} \gtrsim 100 \ {\rm eV}$) of order $\sigma_{\rm brem} \lesssim 10^{-25} \ \rm cm^2$ $(\sigma_{\rm brem}\lesssim 10^{-27} \ \rm cm^2$). Ignoring the effects of elastic and inverse Compton scattering, as well as the possibility of transient ISM phenomena over long time scales, we conservatively estimate a density
\begin{align}
\dot{n}_{E<0} \simeq & \sigma_{\rm brem}(E_{\gamma} \gtrsim E_{\gamma}^{\rm min}) \ n_{E>0}^2 \left(\frac{T}{m_n}\right)^{1/2} \simeq \\ \nonumber & 2 \cdot 10^2 \ {\rm cm^{-3} \ Gyr^{-1}} \left(\frac{n_{\rm ISM}}{10^{-2} \ \rm cm^{-3}}\right) \left(\frac{T_{\rm ISM}}{8000 \ \rm K}\right)^{1/2}
\end{align}
of bound ISM particles that results from the constant loss of energy to photons over $\sim \rm Gyr$ timescales. In the above expression, we assume the relative velocity is dominated by the proton speed within the boson star and, in the rightmost expression, we have used approximate scalings of the bremsstrahlung cross-section that are valid for proton energies $\gtrsim \rm MeV$. Furthermore, we fixed $E_{\gamma}^{\rm min} \simeq \rm eV$ for the numerical estimate above, which yields the maximum correction for boson stars residing in the densest WIM regions. While this number density can be up to order $10^2$ greater than the population of unbound particles, depending on the precise ISM density and temperature, we point out that even accounting for this correction the long mean free path condition set by Eq.~\eqref{eq:mfp-1} is not violated and back-action effects are still negligible for the parameter space under consideration. As mentioned above, the accumulation rate over long timescales requires a careful modeling of elastic interactions between charged particles, and the energy loss/absorption through various inelastic processes. We have additionally verified that the total baryonic mass contained in boson stars at any given time is small compared to the total boson star mass, so the self-gravity of ISM particles can be neglected in our treatment.

\section{Additional Subdominant Radiative Processes}
\label{sec:app-e}
In this Appendix, we comment on other radiative processes that may proceed inside a boson star interacting with the ISM, but are subdominant compared to the emission of final state radiation in collisions. These are associated with the creation of secondaries, and are detailed as follows:
\begin{enumerate}
    \item \textit{Pair Production:} This process has been extensively studied for high energy astrophysical plasmas  \cite{1971SvA....15...17B,1980AcA....30..371Z,1981ApJ...251..713L,1982ApJ...253..842L,1987MNRAS.228..681B,Svensson:1982ia,Svensson:1982hz,1987MNRAS.228..681B,1989ApJ...344..232G}. Particle collisions may produce $e^- e^+$ pairs via the reactions $e^- + p \rightarrow e^- + p + e^- + e^+$ and $e^- + e^- \rightarrow e^- + e^- + e^- + e^+$. In terms of the relative Lorentz factor of the collision, $\gamma_r = (E_1 E_2 - p_1 p_2 \cos\theta)/m_1 m_2$, the threshold for these processes are respectively $\gamma_r \simeq 3$ and $\gamma_r \simeq 7$ \cite{1983MNRAS.202..467S}. If thermalization with the boosted protons is inefficient, electrons in the warm and hot ISM phases will have typical energies of order $\rm eV$ to $\rm 100 \ \rm eV$, which are insufficient for pair production through $e^- e^-$ collisions. On the other hand, although protons can be accelerated to energies in excess of $\sim 100 \ \rm MeV$ for parameters considered, the threshold for the first channel is not attained either. Alternatively, pair production could proceed through hard photon scattering against both electrons and protons. However, these processes will not be dominant so long as the plasma is optically thin, which is the case throughout this work. To illustrate this, both reactions $\gamma + p \rightarrow \gamma + p + e^- + e^+$ and $\gamma + e^- \rightarrow \gamma + e^- + e^- + e^+$ have cross-sections with an asymptotic form $\sigma \sim \sigma_T \times \log(E_\gamma/m_e)$ in the limit $E_{\gamma} \gg m_e$, where $\sigma_T \sim 10^{-25} \ \rm cm^2$ is the Thomson cross-section. For the plasma volume considered, this cross-section is so small that most photons do not scatter within the boson star. In a similar manner, two-photon pair production $\gamma + \gamma \rightarrow e^- + e^+$ has a negligible cross-section as well, of order $\sigma_T$ or below, at the energies considered. Thus, we conclude that pair annihilation either through particle collisions or photoproduction is not relevant in this regime.
    
    \item \textit{Secondary Pion Production:} For moderate energies of order $\sim 300 \ \mev$, $p p$ collisions can produce $\pi^0$ and $\pi^{\pm}$ mesons, through the excitation of intermediate isobar states. The production of secondary pions through this channel has been extensively studied in the context of cosmic rays scattering against cold, neutral hydrogen in the ISM, the so-called hadronic contribution to the gamma-ray spectrum \cite{1968ApJ...151..881S,1970Ap&SS...6..377S,1981Ap&SS..76..213S,1986A&A...157..223D}. Neutral pions contribute directly to the radiation output via the prompt decay $\pi^0 \rightarrow 2\gamma$. On the other hand, the charged pions predominantly decay into $\mu^{\pm}$ ($+\ \nu_\mu \bar{\nu}_\mu$), which subsequently decay into $e^{\pm}$ ($+ \ \nu_e \bar{\nu}_e$), and later annihilate into gammas with a cross-section for the latter of order $\sim \alpha^2 m_e^{-2}$. A reasonable estimate of the overall contribution of these decays can be obtained by utilizing the thermal production rate $\langle \sigma v \rangle_{T}$ tabulated in \cite{1979AZh....56..338K}, and approximating all pions to decay at rest near threshold. In the case of neutral pions, the photon emissivity in this approximation is subsequently given by,
    \begin{equation}
        \left(\frac{d^{3}N_\gamma}{dV dt dE_\gamma}\right)_{\pi^0} \simeq 2 n^2 \langle \sigma v \rangle_{T} \delta(E_\gamma-m_{\pi^0}/2),
    \end{equation}
    where $n$ and $T$ are set by Eqs.~\eqref{eq:ISMaccret1} and \eqref{eq:ISMaccret2}, and $m_{\pi^0} \simeq 134 \ \mev$ is the neutral pion rest mass. For temperatures near the production threshold, $\langle \sigma v \rangle_{T} \simeq 10^{-16} \ \rm cm^{3} \ s^{-1}$, and we find secondary neutral pion production introduces a mild contribution of order $\sim (10$ - $30) \%$ to the bremsstrahlung emissivity at energies $E_\gamma \gtrsim 70 \ \mev$. Indeed, we verified that this estimate is well aligned with a more comprehensive spectrum calculation, utilizing inclusive cross-sections $pp\rightarrow \pi X$ measured at fixed-target experiments and suitable distribution functions for the produced pion in the $p p$ collision \cite{1986ApJ...307...47D}. In a similar manner, the decay of charged pions near threshold will ultimately produce $e^- e^+$ pairs with low-energies that subsequently annihilate. Taking the charged pion decay to also occur at rest, the resulting emissivity from $e^- e^+$ annihilation is negligible compared to the $e^- p$ bremsstrahlung rate at energies $E_\gamma \gtrsim 0.511 \ \mev$.
\end{enumerate}
Based on the above considerations, we find thermal $e^{-}p$ bremsstrahlung is the main contribution to radiation at photon energies $E_\gamma \lesssim 20 \ \rm MeV$, whereas at a sufficiently high temperature, secondary $\pi^0$ production introduces a modest correction to the radiation output at $E_\gamma \gtrsim 70 \ \mev$. Since this correction lies outside the energy range of COMPTEL, we have only considered the former process, with an emissivity given by Eq.~\eqref{eq:bremrate2}. We highlight that thermonuclear reactions could also occur at temperatures $\gtrsim \mev$ \cite{Kafexhiu:2012mx,Kafexhiu:2018qxk,Kafexhiu:2018mzz}. In our simplified description of the ISM composition, the only possible reaction would be $pp\rightarrow D\gamma$, which proceeds through the weak force at a very low rate. To avoid introducing additional uncertainties in our analysis, we have left a complete description of these reactions for future investigation. 

\section{Boson Stars as Point Sources in the Warm and Hot ISM}
\label{sec:app-f}

Here, we analyze the parameter space region where boson stars could be detectable point sources. 
In particular, we consider the Chandra X-Ray Observatory for the hard x-ray end of the spectrum, which is expected to be operational until 2025 and has a point source continuum sensitivity of order $\sim 10^{-15} \ \rm {erg \ cm^{-2} \ s^{-1}}$ \cite{2016ApJ...819...62C}, roughly the same as XMM-Newton's \cite{Brunner:2007kv}. Athena and Lynx, the successors of these missions, are in early stages of development and will not be ready within this decade. 
For soft gamma-rays, on the other hand, we consider the upcoming e-ASTROGAM, which is scheduled to launch in 2029 and will improve sensitivity by about two orders of magnitude compared to COMPTEL, with an estimated value of order $\sim 10^{-12} \ \rm {erg \ cm^{-2} \ s^{-1}}$ assuming reasonable exposure times \cite{2018JHEAp..19....1D}. 

\begin{figure*}[!th]
    \centering
     \includegraphics[width=0.495\linewidth]{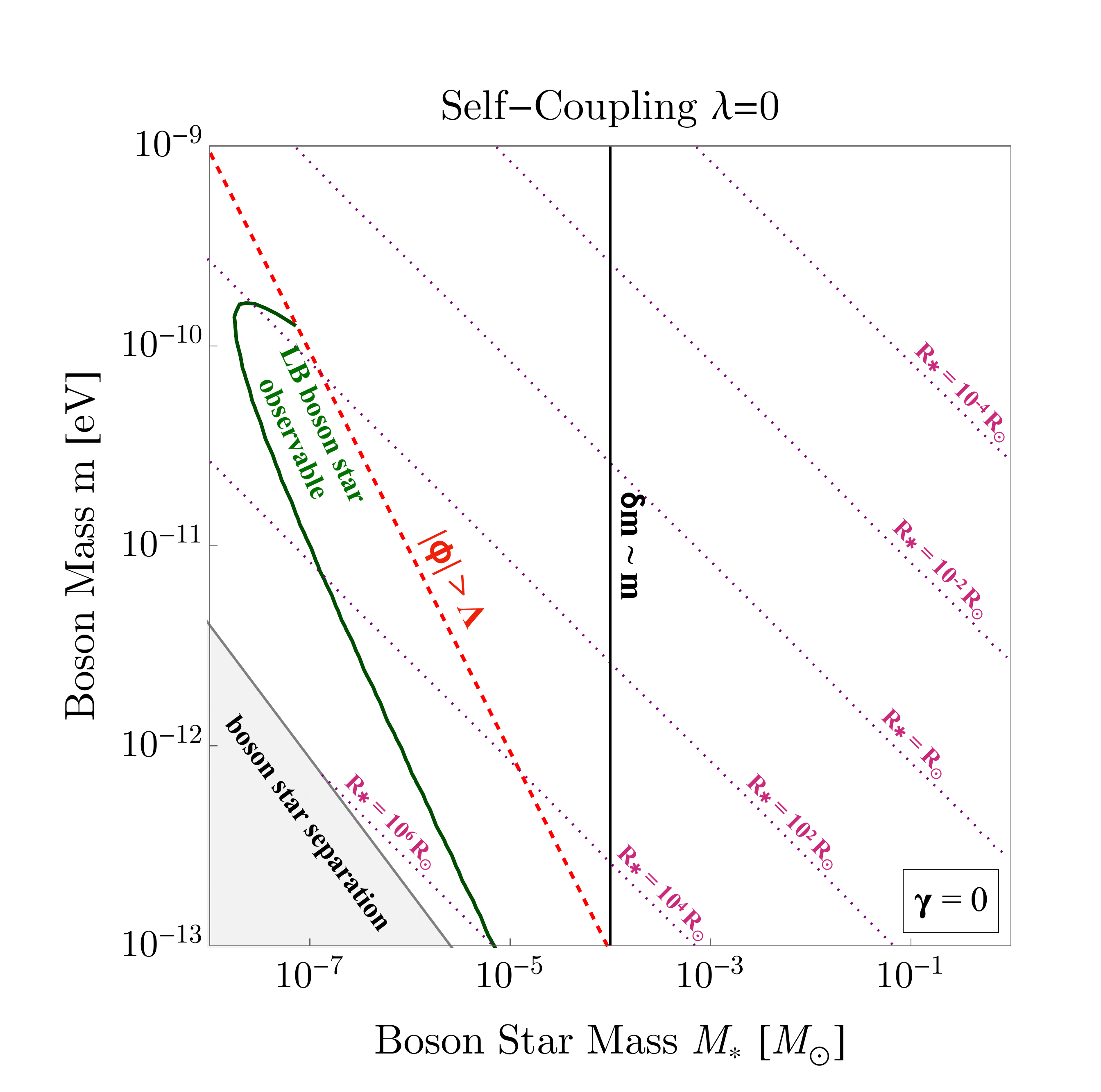}
     \includegraphics[width=0.495\linewidth]{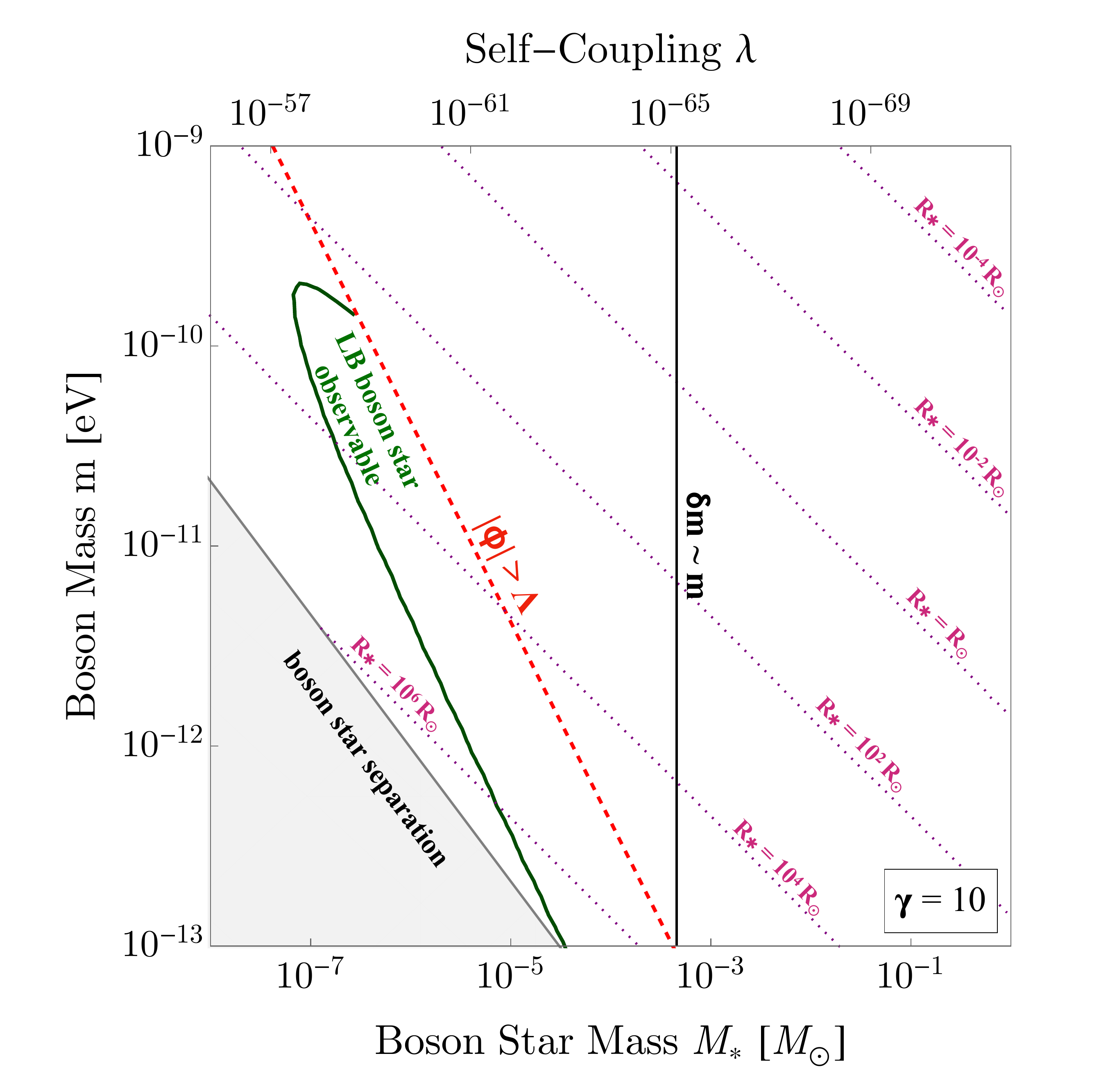}
     \includegraphics[width=0.495\linewidth]{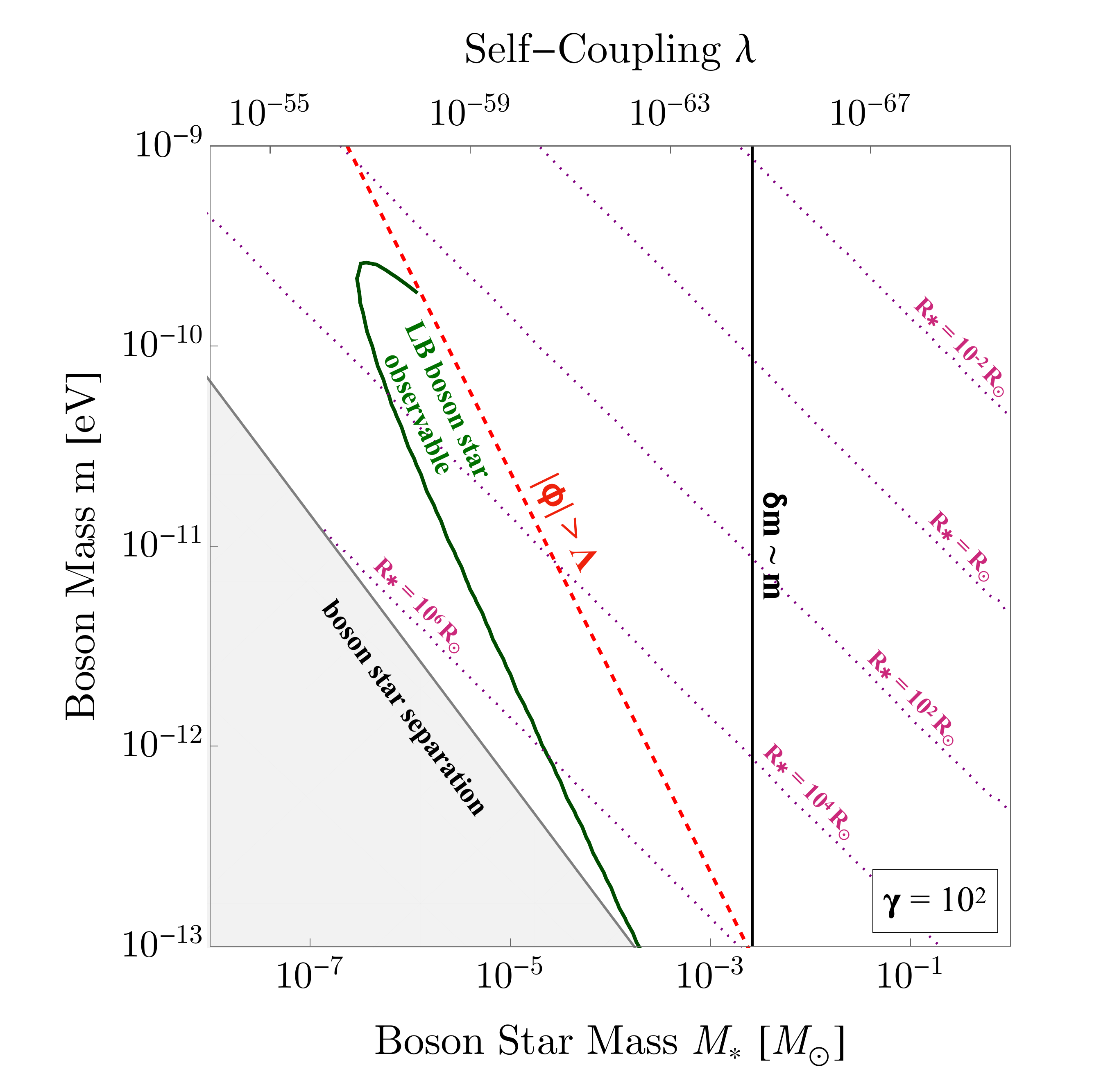}
     \includegraphics[width=0.495\linewidth]{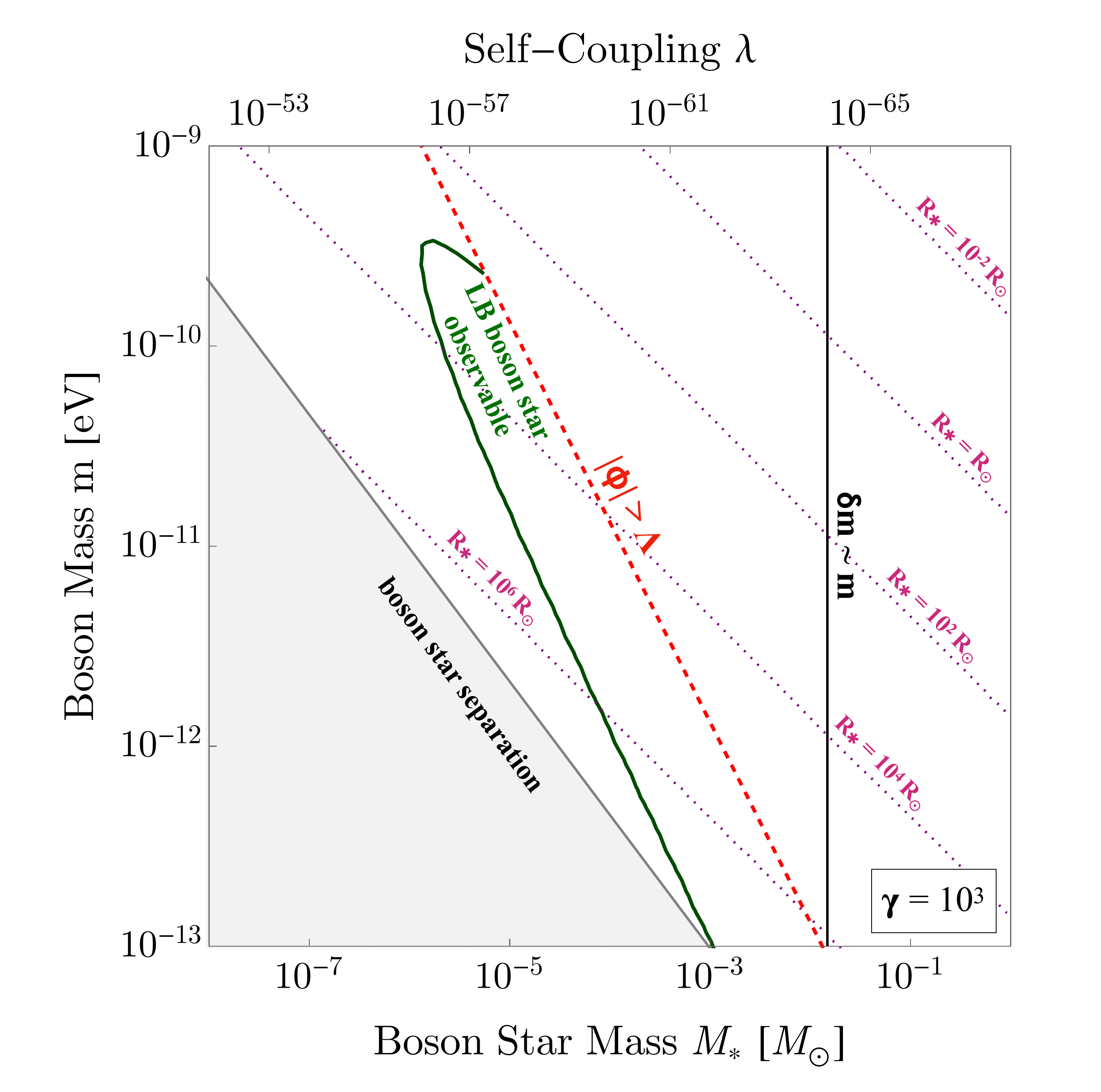}
    \caption{Parameter space where the nearest boson star in the Local Bubble could be observed with the upcoming e-ASTROGAM in the $(1$, $20) \ \rm MeV$ band (\textbf{solid green}), for a fixed boson-nucleon effective coupling $\Lambda = 2 \ \rm TeV$ and various self-interaction strengths as specified in the top axes. For large boson masses, the boson star potential becomes relativistic (\textbf{dashed red}); our analysis is yet to be extended into this regime. The vertical line (\textbf{black}) indicates where back-action effects from ISM particles on the boson star solution become sizable. We also show boson star radius contours (\textbf{dotted purple}), and indicate the parameter space where boson star radii exceed the average separation of these objects in the halo (\textbf{gray}).}
    \label{fig:frac-intcomp1}
\end{figure*}

\begin{figure}[h!]
\centering 
    \centering
     \includegraphics[width=0.495\linewidth]{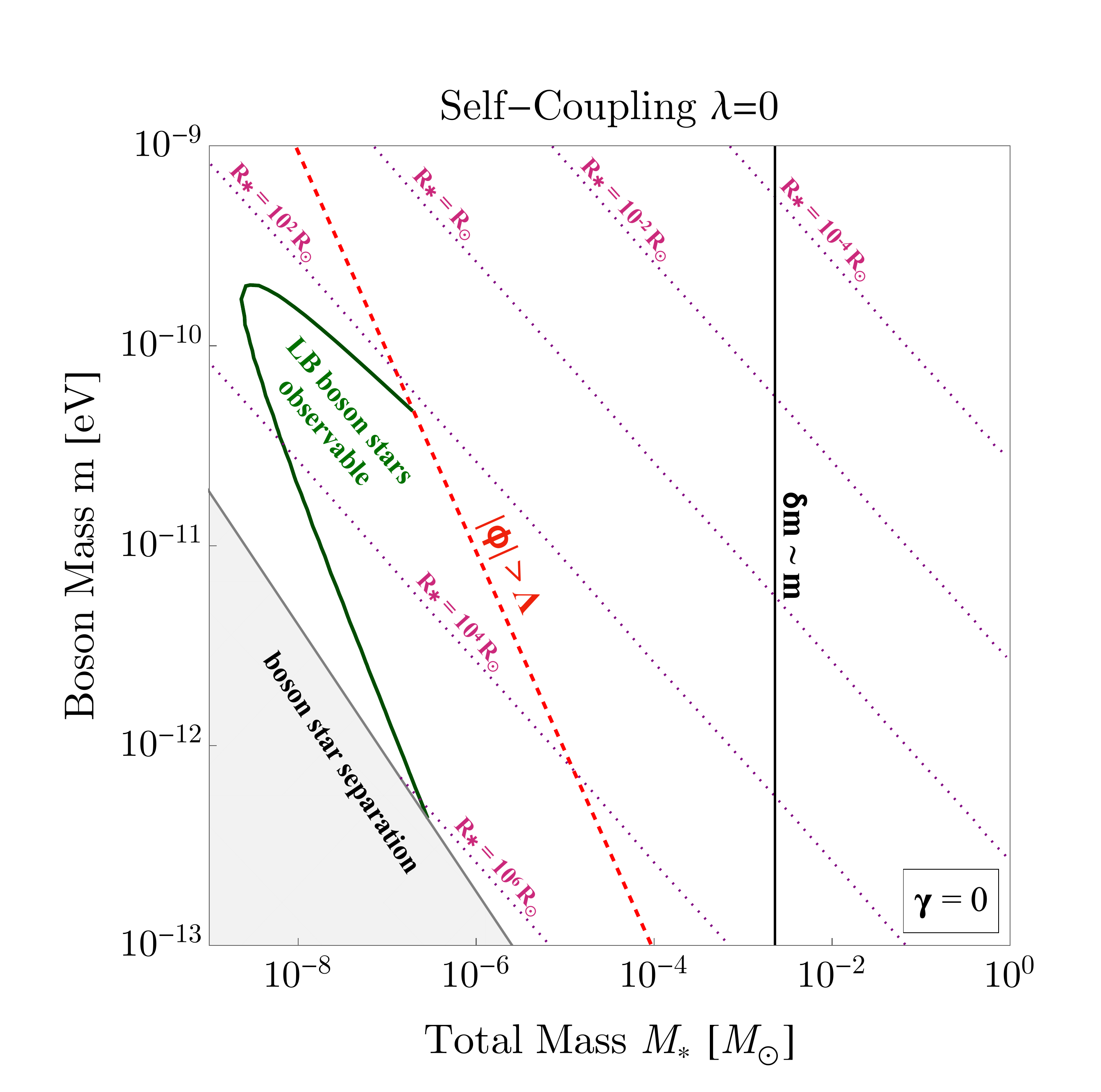}
     \includegraphics[width=0.495\linewidth]{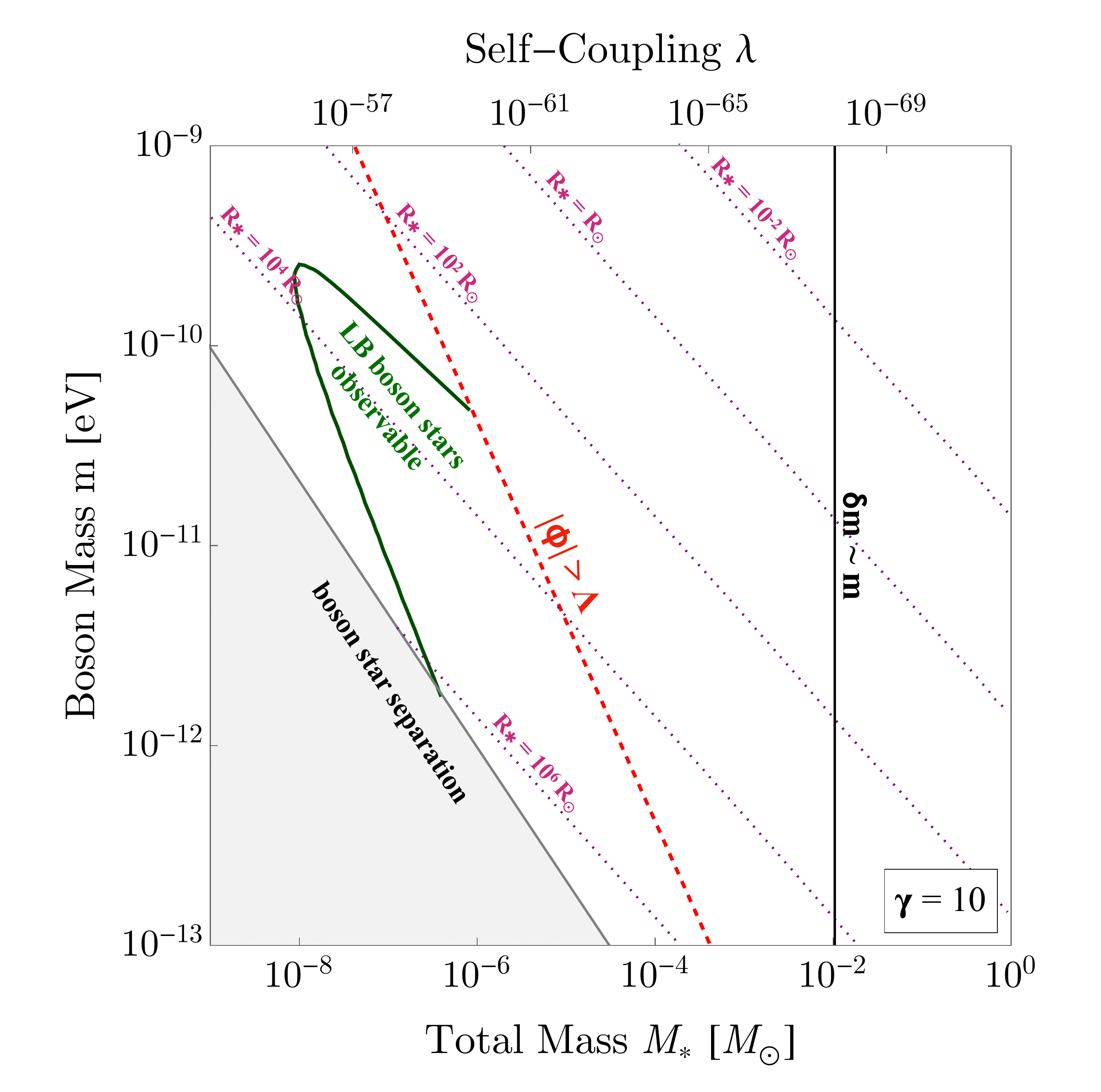}
     \includegraphics[width=0.495\linewidth]{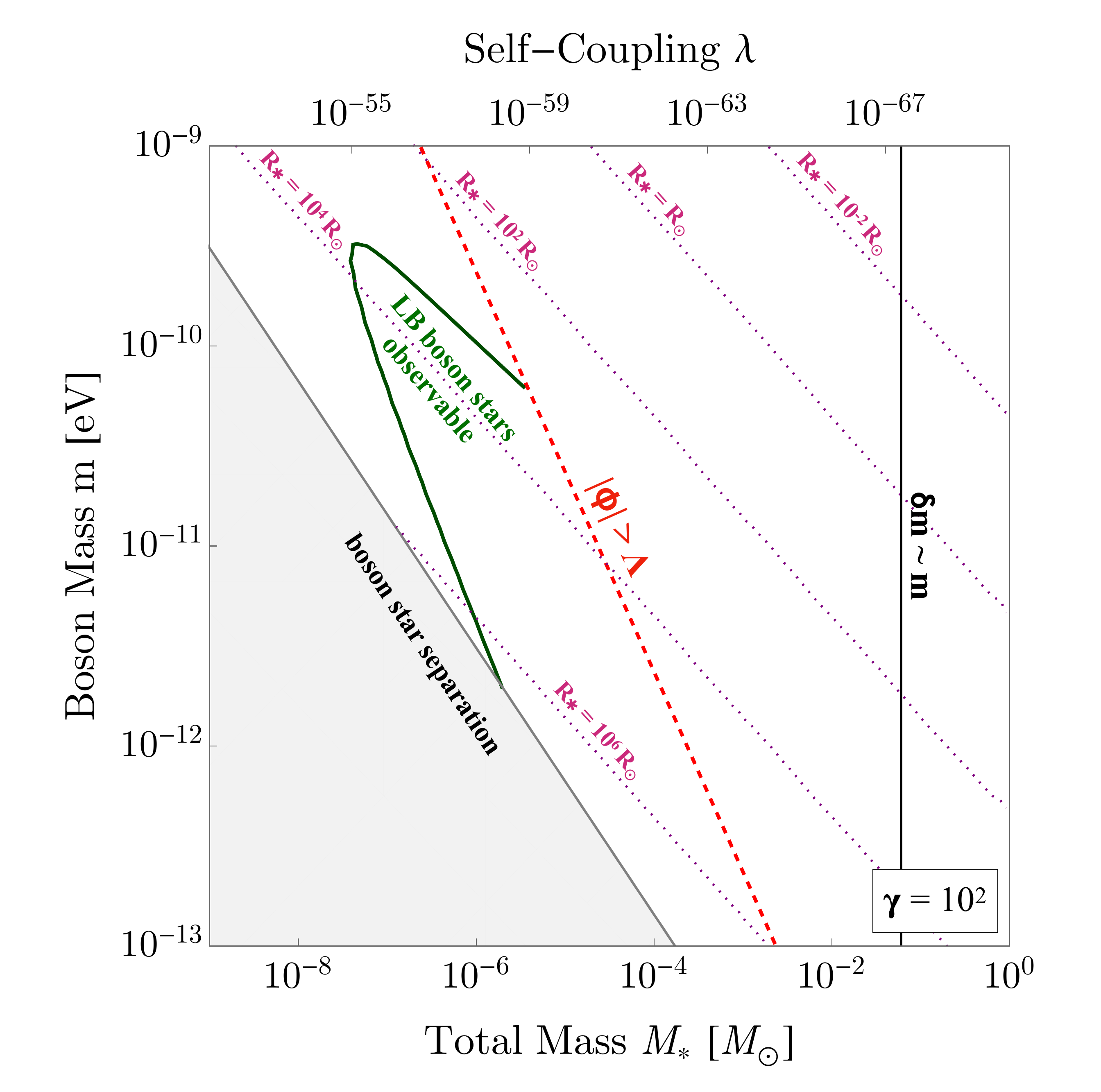}
     \includegraphics[width=0.495\linewidth]{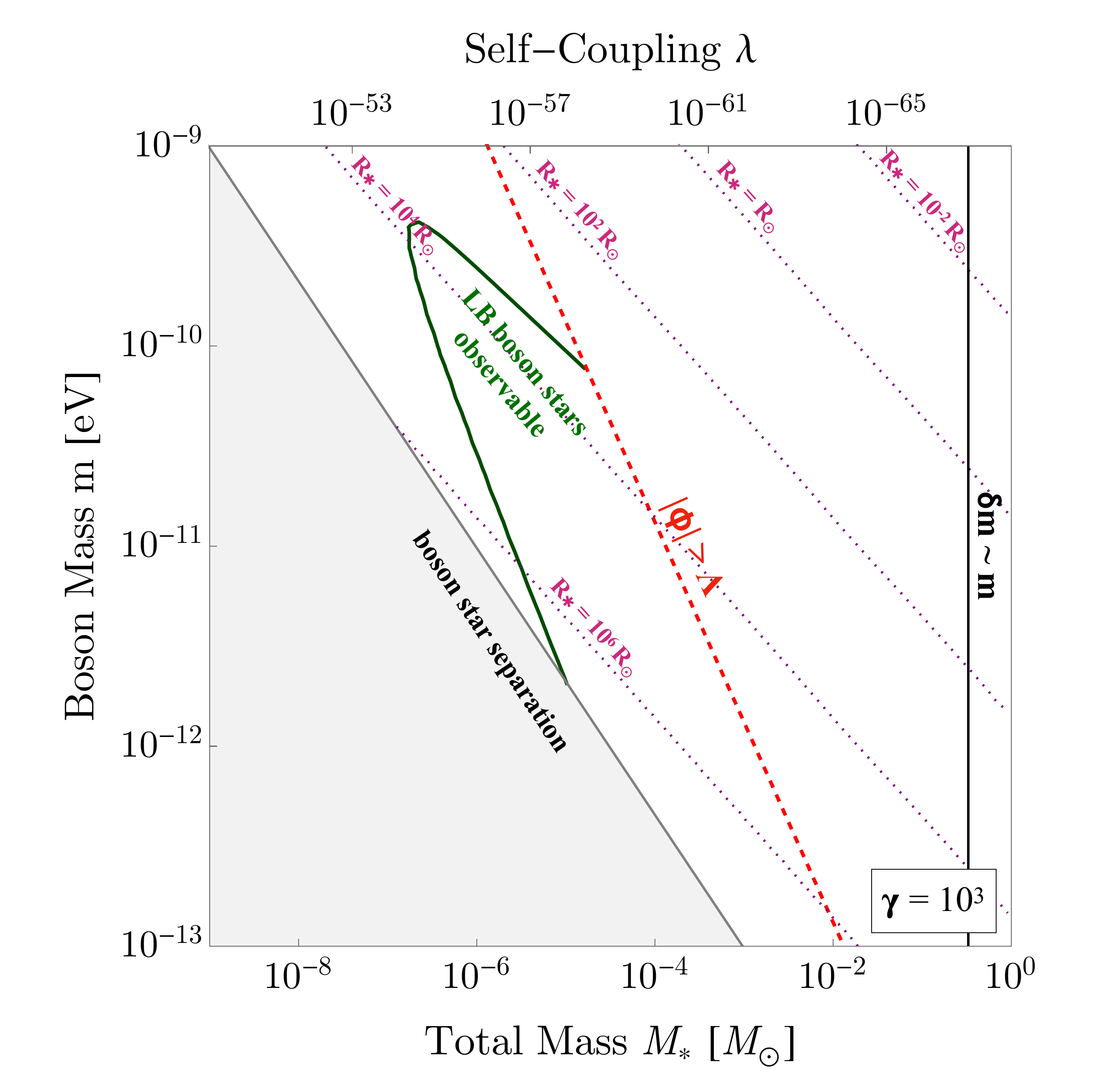}
\caption{Same as Figure~\ref{fig:frac-intcomp1}, but now the parameter space where the nearest boson star in the Local Bubble could be observed with Chandra in the $(1$, $10) \ \rm keV$ band is shown (\textbf{solid green}). As before, we have fixed the boson-nucleon effective coupling to $\Lambda = 2 \ \rm TeV$ and indicated parameter space where the non-relativistic treatment breaks down (\textbf{dashed red}) and back-action effects are not negligible (\textbf{black}).}
\label{fig:frac-heao1}
\end{figure}

Figures~\ref{fig:frac-intcomp1} and \ref{fig:frac-heao1} show enclosed regions of parameter space where the nearest boson star within the Local Bubble ($n_{\rm LB} \simeq 3 \times 10^{-3} \ \rm cm^{-3}$, $T_{\rm LB} \simeq 10^6 \ \rm K$, and $\lesssim 10 \ \rm pc$ radius) is an x-ray/gamma-ray point source with a flux detectable by the above observatories assuming an effective coupling scale of $\Lambda = 2 \ \rm TeV$, marginally above experimental limits. Comparing these regions to the main plots in Figures~\ref{fig:fractionbounds1} and \ref{fig:fractionbounds2}, we see that for $m \lesssim 10^{-10} \ \rm eV$ and couplings to the Standard Model near existing limits, the nearest boson star in the Local Bubble could be accessible to these observatories. Although the direct detection of other boson stars as point sources may depend on the precise details of the local ISM structure, within our global modeling of the ISM ($cf.$ Figure~\ref{fig:ISMstruc}) we find boson stars at distances of order $\gtrsim 50 \ \rm pc$ or greater would be undetectable even with these instruments. In future work, we will analyze point source catalogs and determine what additional constraints on boson stars could be drawn based on the number and observed properties of nearby Galactic point sources. 

\bibliographystyle{JHEP}
\bibliography{cdm}

\end{document}